\documentclass[a4paper,twocolumn,11pt]{quantumarticle}
\pdfoutput=1
\usepackage[utf8]{inputenc}
\usepackage[english]{babel}
\usepackage[T1]{fontenc}
\usepackage[numbers]{natbib}
\usepackage[resetlabels]{multibib}

\usepackage[normalem]{ulem}
\usepackage{subcaption}
\usepackage{makecell} 
\usepackage{graphicx}
\usepackage{dcolumn}
\usepackage{bm}
\usepackage{pythonhighlight}
\usepackage{braket}
\usepackage{array}
\usepackage{pbox}
\usepackage[11pt]{moresize}
\usepackage{mathtools}

\usepackage{multirow, booktabs}
\usepackage{ulem}
\usepackage{listings}
\definecolor{codegreen}{rgb}{0,0.6,0}
\definecolor{codegray}{rgb}{0.5,0.5,0.5}
\definecolor{codepurple}{rgb}{0.58,0,0.82}
\definecolor{backcolour}{rgb}{0.95,0.95,0.92}
\lstdefinestyle{mystyle}{
  backgroundcolor=\color{backcolour}, commentstyle=\color{codegreen},
  keywordstyle=\color{magenta},
  numberstyle=\tiny\color{codegray},
  stringstyle=\color{codepurple},
  basicstyle=\ttfamily\scriptsize,
  breakatwhitespace=false,         
  breaklines=true,                 
  captionpos=b,                    
  keepspaces=true,                 
  numbers=left,                    
  numbersep=5pt,                  
  showspaces=false,                
  showstringspaces=false,
  showtabs=false,                  
  tabsize=2
}

\def\tablename{Table}

\begin{document}


\title{TeD-Q: a tensor network enhanced distributed hybrid quantum machine learning framework}

\author{Yaocheng Chen}
\thanks{Equal contribution; This work was done when
he was a research intern at JD Explore Academy}
\affiliation{Department of Physics, National Taiwan University, Taipei, Taiwan
}
\affiliation{Leung Center for Cosmology and Particle Astrophysics, National Taiwan University, Taipei, Taiwan
}
\affiliation{JD Explore Academy, Beijing, China
}

\author{Chung-Yun Kuo}
\affiliation{Department of Physics, National Taiwan University, Taipei, Taiwan
}
\affiliation{Leung Center for Cosmology and Particle Astrophysics, National Taiwan University, Taipei, Taiwan
}
\affiliation{JD Explore Academy, Beijing, China
}

\author{Yuxuan Du}
\thanks{Corresponding author}
\email{duyuxuan123@gmail.com}
\affiliation{JD Explore Academy, Beijing, China
}

\author{Dacheng Tao}
\thanks{Corresponding author}
\email{dacheng.tao@gmail.com}
\affiliation{JD Explore Academy, Beijing, China
}

\author{Xingyao Wu}
\thanks{Equal contribution; Corresponding author}
\email{wu.x.yao@gmail.com}
\affiliation{JD Explore Academy, Beijing, China
}


\begin{abstract}
TeD-Q is an open-source software framework for quantum machine learning (QML), variational quantum algorithm (VQA), and simulation of quantum computing. It seamlessly integrates classical machine learning libraries with quantum simulators, giving users the ability to leverage the power of classical machine learning while training quantum machine learning models. TeD-Q supports auto-differentiation that provides backpropagation, parameters shift, and finite difference methods to obtain gradients. With tensor contraction, simulation of quantum circuits with large number of qubits is possible. TeD-Q also provides a graphical mode in which the quantum circuit and the training progress can be visualized in real-time.

\end{abstract}

\maketitle


\section{\label{sec:level1}{Introduction}} 

Quantum computing is one of the most promising directions that may save us from Moore's law. Recent results show that quantum computing is indeed advantageous over classical computing in certain problems~\cite{arute2019quantum,madsen2022quantum, wu2021strong}. With the progress in quantum hardware, quantum software and research in algorithms are also catching up. Quantum software plays a crucial role in bridging quantum hardware and real-world applications. There has been a decent amount of work in quantum software platforms in recent years. Some of these already allow users to access real quantum hardware~\cite{bergholm2018pennylane,qiskit,yao}, while others focus more on building a complete software environment for better quantum algorithm development~\cite{cirq}.

TeD-Q is a Python-based quantum programming framework that has differentiable functionality. It is optimized particularly for quantum machine learning problems~\cite{biamonte2017quantum,schuld2015introduction,schuld2019quantum,huang2021power} and variational quantum algorithms~\cite{yuan2019theory,cerezo2021variational,bittel2021training}, which constitute the majority of contemporary Noisy Intermediate-Scale Quantum (NISQ) algorithms~\cite{preskill2018quantum,bharti2022noisy}. TeD-Q provides a universal framework for programming on different backends, including quantum hardware, quantum simulator, and distributed GPU accelerators. In terms of top-level applications, users could enjoy the convenience of using TeD-Q without worrying about how to deal with different backends. One could simply treat a quantum neural network (QNN) training as in classical machine learning. 

With TeD-Q, the quantum circuit is treated as a Python function. It is free to choose either PyTorch or JAX as the interface with the execution backends. Due to this systematic integration, one could also leverage the rich features provided by the well-developed AI-oriented libraries~\cite{NEURIPS2019_9015} to help the implementation of quantum algorithms better. Specific benefits include batch execution, Just In Time compilation (JIT)~\cite{aycock2003brief}, distributed parallel GPU optimization, and auto-differentiation for backpropagation.

TeD-Q has both full amplitude quantum simulation and a tensor network enhanced multi-amplitude simulation. TeD-Q will estimate the complexity of both modes, and users could choose on their own which mode to execute the code. 

TeD-Q is also equipped with a built-in tensor network module and contraction path optimizer JDtensorPath, which could provide additional improvements on the performance in the tensor network simulation mode. This is achieved by optimizing the tensor network contraction order.

Besides, TeD-Q provides a graphical circuit composer that makes it user-friendly to beginners with intuitive visualization.

Table~\ref{Tab:platforms} shows the comparison among TeD-Q and three major general-purpose quantum machine learning platforms -- PennyLane~\cite{bergholm2018pennylane}, Qiskit~\cite{qiskit} and Paddle Quantum~\cite{Paddlequantum}. 

PennyLane is a cross-platform Python package for programming quantum and hybrid quantum-classical algorithms. It builds upon the idea of auto differentiable programming and integrates mainstream classical machine learning libraries with quantum algorithms to provide an easy-to-use interface\cite{bergholm2018pennylane}. Qiskit is an open source quantum software development framework of IBM. It integrates PyTorch and can be executed with IBMQ on real quantum hardware\cite{qiskit}. Paddle Quantum is developed by Baidu, it is based on the PaddlePaddle deep learning platform and can access real quantum computers Quantum Leaf\cite{Paddlequantum}.

\begin{table}[h!]
	\centering
	\caption{A summary of features of existing quantum simulation frameworks.}
	\ssmall
    \begin{tabular}{c|c|c|c|c}
          & TeD-Q & Pennylane & Qiskit & \makecell{Paddle\\Quantum} \\
        \Xhline{1.2pt}
        Backpropagation  & v &  v  & - & v  \\
        \hline
        Hypergraph-based TNC & v & -  &  - & -  \\
        \hline
        Slicing paralleling  & v &  -  & - & -  \\
        \hline
        PyTorch interface  & v &  v  & v & -  \\
        \hline
        Noise model  & future release &  v  & v & v  \\
        \hline
        Hardware compatibility & \pbox{2cm}{via Qiskit,\\ \centering Quafu} & v & v & v  \\
        \hline
        Circuit composer & v & - & v & -
    \end{tabular}
    \label{Tab:platforms}
\vspace{-1em}
\end{table}

The main distinct features of TeD-Q are hypergraph-based tensor network contraction (TNC) and paralleling via index slicing. Simulation of quantum circuits normally can be carried out by matrix multiplication and tensor network contraction. Current quantum software platforms generally support only matrix multiplication mode~\cite{bergholm2018pennylane,qiskit,Paddlequantum}, which is enough while the number of qubits is below roughly 38. However, when the number of qubits exceeds 38, the time cost and storage cost surge exponentially. The TNC mode of TeD-Q could easily handle quantum circuit simulation with qubit number larger than 38. Hypergraph-based TNC method can provide several orders of magnitude improvement in both computation and memory complexity while paralleling via index slicing delivers advantages in GPU device communication. Since the order of tensor network contraction will affect the time cost a lot, TeD-Q also provides a dedicated tensor network contraction optimizer, called JDTensorPath.  These new features are designed to improve the efficiency of large-scale quantum circuit simulation.

Compared to Qiskit, Pennylane and Paddle Quantum, TeD-Q also featured a reusable mechanism. In most of the quantum algorithms including NISQ algorithms, the parameterized quantum circuit will be evaluated multiple times without changing the circuit structure. In TeD-Q, the quantum state and quantum gate objects inside a quantum circuit will be reused in every evaluation. This feature can save a considerable amount of time for class instantiation in small quantum circuits. This advantage could be easily seen in the performance comparison of multiple qubits rotation example in Section~\ref{Performance}.

\section{Architecture of TeD-Q} 
\label{archi}
\begin{figure}[!h]
    \centering
    \includegraphics[width=0.48\textwidth]{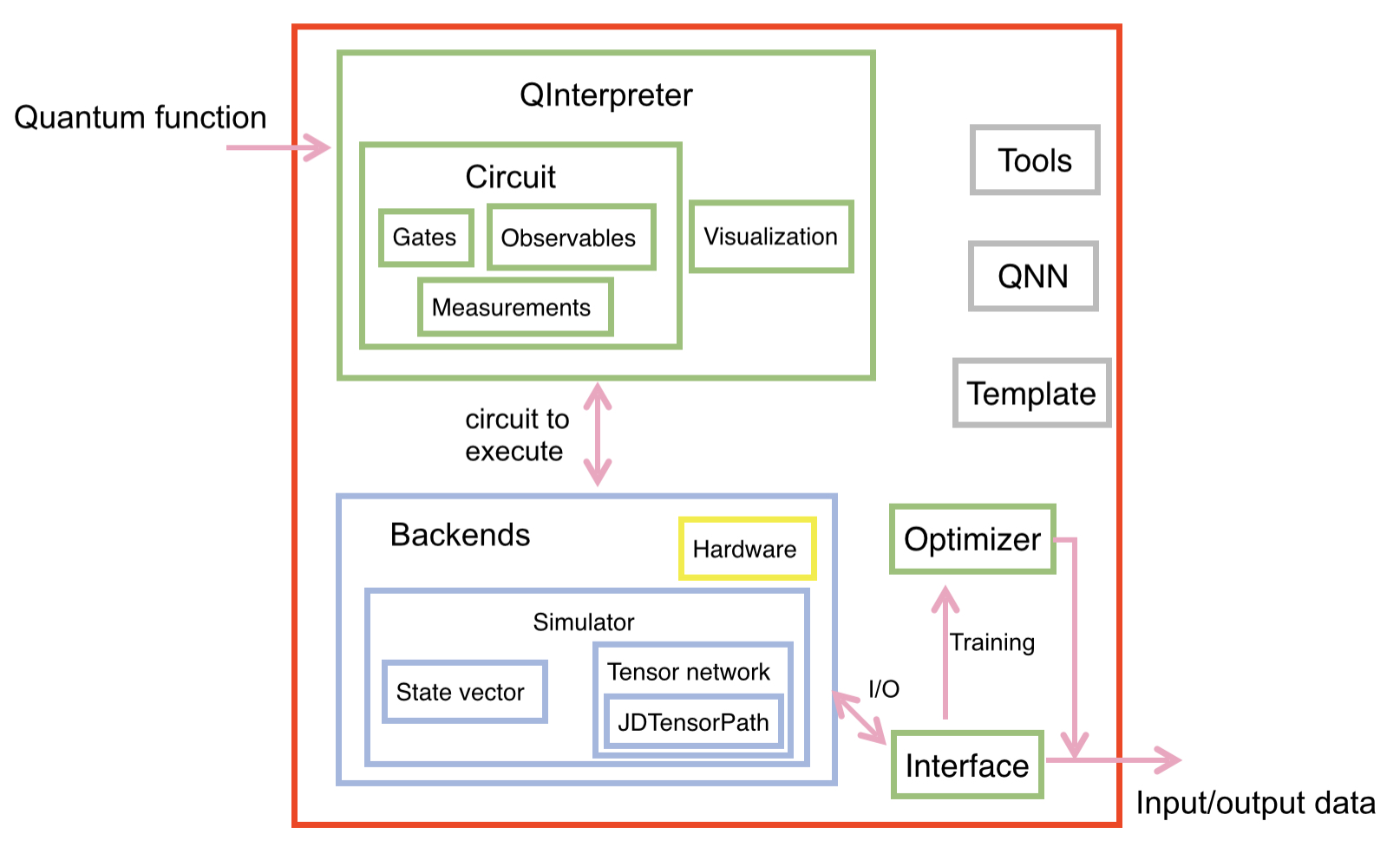}
    \caption{The design of TeD-Q architecture. The part in yellow is quantum while the rest is classical processing.} 
    \label{architecture}
\end{figure}

TeD-Q is designed for fast and efficient implementation of quantum algorithms. For this purpose, the components of application design, computation backends, and the interface in between are packed into relatively independent modules. TeD-Q is composed of four main modules, QInterpreter, Backends, Interface, and Optimizer, as shown in Fig.~\ref{architecture}. 

Any quantum algorithm will eventually be passed through and processed as classical codes. This is where the QInterpreter plays its role, transforming quantum algorithms into executable codes. Quantum state and quantum gate classes are the basic building blocks of QInterpreter. A series of quantum gates applied to the quantum state will make up a quantum circuit.  

The quantum circuit object in TeD-Q is agnostic with respect to simulation and hardware backends. It is an abstraction of user-defined quantum function, that bridges the high-level quantum algorithms and low-level backend-specific instructions. QInterpreter of TeD-Q provides tools for obtaining and manipulating representations of quantum operators and measurements. These tools and representations are useful for generating a quantum circuit. The user needs to provide a quantum function, the number of qubits, and the parameters or the shape of the parameters (in this case, random parameters will be generated and used) to construct a quantum circuit.

A quantum function is defined as a normal Python function. It accepts classical input parameters and uses their elements for quantum gate parameters. The user needs to put quantum operations inside the quantum function in the order that matches how they are applied. A keyword argument "qubit", which is a list of integers, need to be specified to denote which qubit each quantum operation applies on. After all the quantum operations, a single or a list of measurements must be placed as part of the return statement. Code Listing~\ref{how_to_construct_a_qc} shows a simple example of defining a quantum function and constructing a quantum circuit from it.

\begin{lstlisting}[caption={How to construct a quantum circuit. Line 4-9 define a quantum circuit function while line 15 constructs a TeD-Q quantum circuit object.}, captionpos=b, label={how_to_construct_a_qc}, language={Python}]
import tedq

# Define quantum circuit
def circuitDef(params):
    tedq.RX(params[0], qubits=[0])
    tedq.RY(params[1], qubits=[1])
    tedq.CNOT(qubits=[0, 1])
    return [tedq.expval(tedq.PauliZ(qubits=[0]))
            tedq.expval(tedq.PauliZ(qubits=[1]))]
            
number_of_qubits = 2
parameter_shapes = [(2,)]

# Quantum circuit construction
circuit = tedq.Circuit(circuitDef, number_of_qubits, parameter_shapes = parameter_shapes)
\end{lstlisting}

The quantum circuit is not yet executable since the computation mode and backend have not been chosen. After specifying these, it can be compiled into an executable compiled circuit. The compiled circuit can be used like a standard Python function to compute the result. The order of input parameters must be the same as the order of the corresponding quantum gates. A quantum circuit can be run on different backends by manually compiling it with different specifications, as shown in Code Listing~\ref{compiled_circuit}. This is why it is called a reusable quantum circuit.

\begin{lstlisting}[caption={Compile quantum circuit with different backends.}, captionpos=b, label={compiled_circuit}, language={Python}]
# compile quantum circuit with "jax" backend
compileCir1 = circuit.compilecircuit(backend="jax")

# compile quantum circuit with "pytorch" backend
compileCir2 = circuit.compilecircuit(backend="pytorch")
\end{lstlisting}

\section{Simulation modes}
\label{simulation_mode}
\subsection{State vector propagation mode} 
In quantum theory, the wave function is used as a mathematical description of the quantum state of a system. A state vector is denoted by a complex ket vector $|\psi\rangle$. In this mode, quantum gates act on the state vector as a sequence of matrix operators, so that the resulting state after each operation is:
\begin{equation}
    |\psi\rangle = U|\psi_0\rangle,
\end{equation}
where $U$ is the matrix operator. The state vector propagation mode is the default simulation backend of TeD-Q. A user-defined initial quantum state can be used by putting tedq.InitStateVector() as the first line of the quantum circuit definition function as shown in Code Listing~\ref{init_state}. Notice that the length of the initial state vector must match $2^n$, where $n$ is the number of qubits. If no initial state is specified, TeD-Q will set it to the ground state $|0\rangle^{\otimes n}$ by default. With TeD-Q, this mode allows a normal laptop to deal with a non-trival quantum circuit of up to $20$ qubits.

\begin{lstlisting}[caption={Define quantum circuit with user-defined initial quantum state.}, captionpos=b, label={init_state}, language={Python}]
# Prepare quantum state vector
init_state = np.array([1./np.sqrt(2.), 1./np.sqrt(2.)])

# Define the quantum circuit with a specific initial quantum state
def circuitDef(parameters):
    tedq.InitStateVector(init_state)
    # other circuit definition
\end{lstlisting}

\subsection{Tensor network contraction mode} \label{section_tnc}
In the state vector propagation mode, the number of amplitudes grows exponentially with the number of qubits, $2^n$ complex numbers are needed to describe the quantum state for an n-qubit system \cite{qiqcbook}. Therefor, it is extremely difficult for classical state vector-based simulator to handle NISQ devices with more than 50 qubits. TeD-Q integrates the tensor network contraction method and mainstream deep learning frameworks -- PyTorch and JAX, providing an efficient built-in tensor network-based quantum simulator. This mode can easily simulate certain quantum circuits of $50 \sim 100$ qubits on a single CPU or GPU.

The workflow of the tensor contraction mode in TeD-Q consists of the following steps: (a) manipulating the quantum circuit according to its output type and converting it into a tensor network by the built-in TensorNetwork module; (b) applying structural simplification on the tensor network; (c) searching for best contraction sequences for the simplified tensor network; (d) slicing the contractions down to memory limit; (e) carrying out the actual contraction with the backend library according to sliced contraction sequences. The computational complexity of steps (a) and (b) are very low, which can be done by a single CPU thread. Step (b) is optional, however, it will generally reduce the size of the tensor network significantly. Except for the default trivial pathfinder, two cutting-edge hyper-graph partition-based packages -- CoTenGra\cite{gray2021hyper} \& JDtensorPath can also be chosen for step (c). JDtensorPath was originally a built-in module in TeD-Q and was separated into an independent software for better code structure management. It perfectly matches TeD-Q's design paradigm and will be introduced in the Section~\ref{section_hyper_graph}. Step (d) can also be carried out with CoTenGra or JDtensorPath libraries. This is also an optional step and will cost some extra overhead of the total contraction floating point operations (FLOPs). In JDtensorPath, step (c) and (d) are done parallelly in CPUs. The path found in step (c) or (d) can be re-used in step (e) to obtain the result of the same quantum circuit but with different input parameters. Step (e) is the most time-consuming part, which will be executed repeatedly in the machine learning training loop. Step (e) can be run on a single CPU, single GPU, or a cluster of GPUs.

\subsubsection{Tensor network and simplification} 

\begin{figure}[!h]
    \centering
    \includegraphics[width=0.36\textwidth]{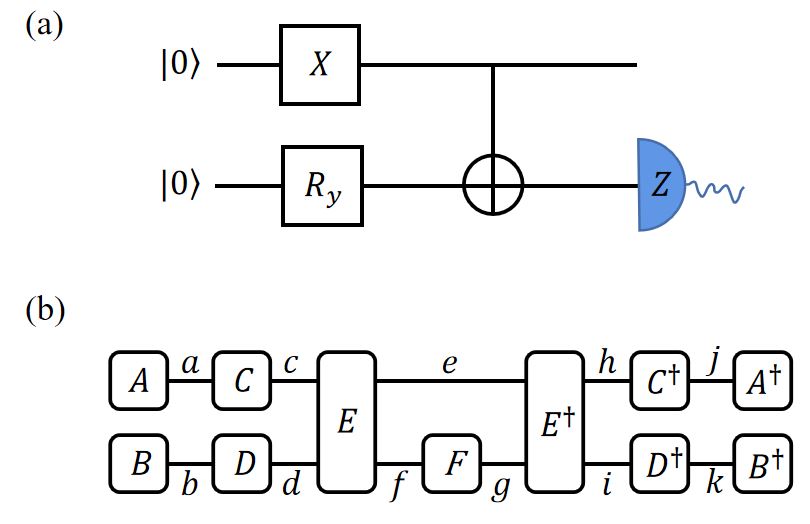}
    \caption{A quantum circuit and its graphical tensor network representation in TeD-Q. (a) A two qubits quantum circuit with expectation value measurement on second qubit. (b) The corresponding graphical tensor network representation, each vertex is associated with a tensor given by quantum state or operator (gate, state or observable).
} 
    \label{circuit_to_tn}
\end{figure}

A quantum circuit can be simulated in various ways, of which tensor network method is increasingly popular recently years. TeD-Q can represent a quantum circuit with the corresponding graphical tensor network. Both the quantum state and any quantum gate can be represented as tensors while the tensor network preserves the topological structure of them. The resulting graph $G = (V, E)$ thus contains vertexes associated with the tensors, while the edges are labeled by their indices. The rank of a tensor is given by the number of edges connecting to it. Since a single qubit lives in a 2D Hilbert space, meaning each index takes value from $\{0, 1\}$, a rank-k tensor will require $O(2^{k})$ storage space. To calculate the probability of finding certain final state $\ket{\phi_{\text{out}}}$ from the output of the quantum circuit $U\ket{\phi_{\text{in}}}$, i.e. $\bra{\phi_{\text{in}}}U\ket{\phi_{\text{out}}}$, the tensor network needs to be contracted, meaning shared indices between vertices will be summed up; while open edges remains.  Fig.~\ref{circuit_to_tn} gives an example of a trivial two qubits quantum circuit and its graphical tensor network representation in TeD-Q. Each vertex in the tensor network is associated with a tensor representing either the quantum state or one quantum gate. $A_{a}$ and $B_{b}$ are tensors corresponding to state $\ket{0}$s of the two qubits while $C_{ca}$ and $D_{db}$ are representing single qubit Pauli $X$ and rotational $R_y$ quantum gates respectively; the two qubits CNOT gate corresponds to $E_{efcd}$ and the measurement in the $z$ direction is presented by $F_{gf}$. Since the expectation value is derived by $\langle I\otimes Z \rangle = \langle00|U^{\dag}(\theta)I \otimes Z U(\theta)|00\rangle$, to calculate it with tensor network, we must append the tensor network corresponding to the complex conjugate of the original quantum circuit, which is $\bra{00}U^\dagger(\theta)$. The $R_y(\theta)$ gate could then be represented by $D_{db}$ which in tensor form will be:

\begin{equation}
D_{db}=
    \begin{bmatrix}
    \cos(\theta/2)&-\sin(\theta/2)\\
    \sin(\theta/2)&\cos(\theta/2)
    \end{bmatrix};
\end{equation}
while the tensor form of the CNOT gate is:
\begin{equation}
\begin{split}
E_{00cd}&=
\begin{bmatrix}
1&0\\
0&1
\end{bmatrix},
E_{11cd}=
\begin{bmatrix}
0&1\\
1&0
\end{bmatrix},\\
E_{01cd}&=E_{10cd}=
\begin{bmatrix}
0&0\\
0&0
\end{bmatrix}.
\end{split}
\end{equation}

Thus, the expectation value can be calculated by the summation:
\begin{equation}
    \begin{split}
        T &= \sum_{a,...,k} A_{a}B_{b}C_{ca}D_{db}E_{efcd}F_{gf}E^*_{hieg}C^*_{jh}D^*_{ki}A^*_{j}B^*_{k}\nonumber\\
        &= \sin^2(\theta/2) - \cos^2(\theta/2),
    \end{split}
\end{equation}
which can be verified with the traditional vector mode calculation. It is shown that the actual order of carrying these summation (for instance, whether to sum $a$ or $k$ first) does matter a lot and finding a good contraction order composes most part of the tensor network algorithm.

TeD-Q supports an efficient local processing of the tensor network prior to the searching of the contraction order by a set of simplifications based on its structure and sparsity of the tensors. The simplifications include tensor shape squeezing, diagonal and anti-diagonal reduction, and rank simplification, which are designed to decrease the complexity and the rank of the tensor~\cite{gray2021hyper}. After the local pre-procession, the tensor network will be transferred from a line graph into a hyper-graph (a generalization of the graph that allows an edge connecting any number of vertices).

\begin{figure}[!h]
    \centering
    \includegraphics[width=0.48\textwidth]{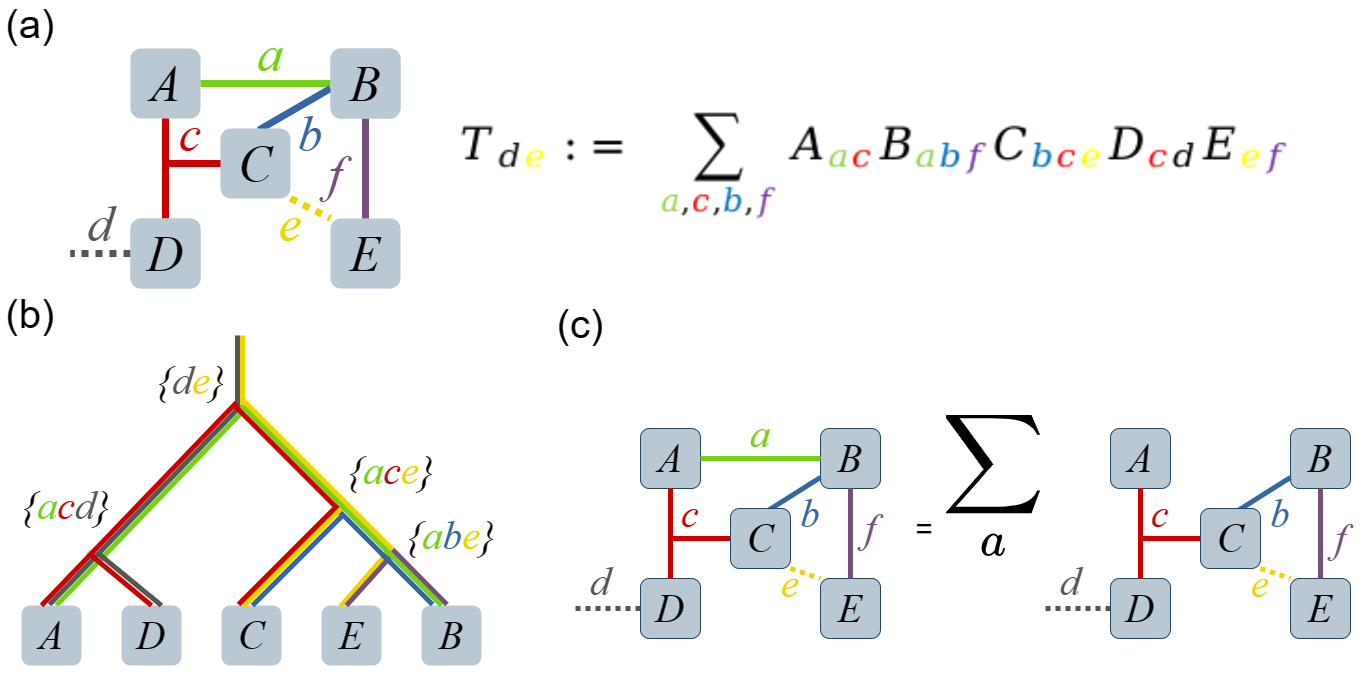}
    \caption{An illustration of tensor network and its hypergraph representation, binary contraction tree, and index slicing. (a) An example of a tensor network and its hypergraph representation, edge $c$ is a hyper-edge connecting three vertices. Dashed lines represent open edges while solid lines correspond to closed edges. (b) A binary contraction tree, each node represents a pairwise contraction of two tensors as well as a bi-partitioning of its hyper-graph. The upward edges indicate the indices of the intermediate tensor. (c) Slicing the green edge (index $a$), generating simpler tensor networks with the same  structure labeled by all possible values of $a$. Each individual contraction is to be carried out independently and the summed of them gives the final result.} 
    \label{tensornetwork}
\end{figure}

An illustration of the tensor network and its hypergraph representation is shown in Fig.~\ref{tensornetwork} (a). Notice that the red hyper-edge (corresponding to index $c$) is connected with three vertices (tensors). The original tensor network is defined as:
\begin{equation}
    T_{de} := \sum_{a,c,b,f} A_{ac}B_{abf}C_{bce}D_{cd}E_{ef},
\end{equation}
where the upper-case and lower-case letters denote tensors and its indices respectively.

\subsubsection{hyper-graph-based tensor contraction path optimizer} \label{section_hyper_graph}
The time and memory complexity of tensor network contraction depend on its tensor ranks and contraction order. Despite finding optimal contraction sequences is theoretically a non-deterministic polynomial-time (NP)-hard task for general cases, efficient heuristic algorithms~\cite{gogate2012complete} exist in practice for obtaining 'good enough' sequences. Several established algorithms are supported by TeD-Q from third party libraries, including brute-force search, branching strategy, greedy and random-greedy path, dynamic programming approach, line-graph tree decomposition via QuickBB or FlowCutter, community detection, Boltzmann-greedy agglomerative contraction trees, and hypergraph partitioning method~\cite{daniel2018opt,gray2021hyper}. The default basic pathfinder of TeD-Q supports the first four methods. For better performance, we adopt the hyper-graph partitioning method which is original proposed in CoTenGra and develop our own hypergraph based pathfinder library -- JDtensorPath. 

The key idea of this algorithm is to recursively divide the hypergraph of a given tensor network, and explicitly build an incomplete binary contraction tree in a 'top-down' fashion as shown in Fig.~\ref{tensornetwork}b. The root node of the tree corresponds to the initial whole graph of the tensor network and the effective tensor of the final result. Bi-partitioning the hypergraph of the parent node will create two child nodes that represent the effective intermediate tensors. The contraction tree can be generated in this recursively divisive manner. The contraction sequences are obtained by ascending the tree from the bottom up, and combining two child nodes to a parent node corresponds to a pair contraction of two intermediate tensors. 

The FLOPs for an individual contraction (node bi-partitioning) is given by:
\begin{equation}
    \text{FLOPs} = p*(2*q-1),
\end{equation}
where $p$ is the product of the dimensions of the outer indices (indices that are not contracted) and $q$ is that of the contracted indices (which equals to $1$ if no index is contracted). For each node bi-partitioning, $p$ is fixed; therefore, the cost of the corresponding pairwise contraction can be minimized by minimizing the sum of the weighted edges (indices) cut by the partition. This method is like a greedy approach from top to down since it only considers the cost of every single contraction. Furthermore, JDtensorPath not only requires partitioning of the current node to be optimal, but also takes the partitioning of its child nodes into consideration, so that the overall tree size and total contraction FLOPs could be minimized.  Specifically, it is required that the outer indices of the parent node must be distributed as evenly as possible on two child nodes. This requirement can reduce $p$ in the partitioning of the child nodes and thus affects the FLOPs of their contractions. The memory usage of a pairwise contraction is given by the sum of the sizes of all the related tensors. Since the parent node tensor of each contraction is fixed, evenly distributing its indices on children node tensors can also reduce the memory usage. This is due to the fact that, supposing $p$ is fixed and defining $m$, $n$ to be the product of the dimensions of remaining indices of children node tensors respectively during contraction, the memory cost of the contraction $m*q + n*q$ will be minimized when $m=n$ given $m*n=p$.

Advanced multilevel hypergraph partitioning framework KaHyPar\cite{DBLP:phd/dnb/Schlag20,ahss2017alenex} is employed to construct the contraction tree, modern parallel and distributed Python package--Ray is used to generating multiple trials in parallel. JDtensorPath can provide high-quality contraction sequences to TeD-Q and enables tensor network contraction mode to achieve orders of magnitude speedup for various quantum circuits simulation.

\subsubsection{Slicing} 
\label{slicing}
Slicing is an optional step. It is designed to solve the memory limit problem and provides an efficient way for parallel computing\cite{huang2021efficient}. Even with an excellent contraction path, contraction of a large tensor network is still limited by its intrinsic memory complexity. In addition, the elaborated data dependency between the tensors becomes a bottleneck for contraction with multiple computational units since it needs a lot of inter-processor communication. Slicing the indices can split the original tensor network into multiple tensor networks with identical structures (same hypergraph), each with a memory cost small enough to fit into a single processing unit. The simulator can execute these sub-tasks in parallel without dependencies or sequentially depending on the computational resources. The value of the original tensor network is given by summing up the partial results obtained from the contractions of each sub-tasks, in which the assignment of the selected sliced indices is fixed\cite{pednault2017pareto}. Take Fig.~\ref{tensornetwork} as an example, after slicing, the original tensor network can be further represented as:
\begin{equation}
    T_{de} := \sum_{i=a} (\sum_{c,b,f} A^{(i)}_{c}B^{(i)}_{bf}C_{bce}D_{cd}E_{ef}).
\end{equation}

Finding the optimal subset of indices to slice is also an NP-hard task with complexity $O(m^n)$, where $m$ and $n$ are the number of all and sliced indices, respectively. The strategy here is to use a heuristically greedy algorithm described as the following: \\
(a) find out all intermediate tensors that are larger than the target size;\\
(b) pick up one index shared by the largest tensors or appears most frequently among step (a) tensors random; \\
(c) repeat the above steps until the contraction width is low enough.

Finding slicing indices is strongly entangled with searching for the best contraction path. Slicing is based on an existing contraction path, but the optimal path for the original tensor network is not guaranteed to be suitable for slicing. Therefore, JDtensorPath proposes a two-phase indices-slicing-incorporated contraction path finding heuristic. The first phase generates contraction tree trials followed by 128 pseudo-slicing trials, which only calculate the FLOPs without actually slicing indices on contraction trees. In the second phase, apply ``real" slicing trials (default is 1024 times) on the contraction tree with the lowest FLOPs after pseudo-slicing. Then, the simulator can obtain a ``good enough" slicing indices and contraction path for sub-tasks in a reasonable searching time. The implementation of the slicing heuristic is based on set operations, and the simulator will store only the best-sliced contraction tree. By doing so, it becomes a very efficient and memory-saving algorithm. Besides, this slicing subroutine can also be carried out in parallel in the CPU computation using the Ray framework.

\subsubsection{Users API} 
TeD-Q enables users to customize their control of the tensor network contraction process. Code Listing~\ref{tn_mode} shows an example of compiling a quantum circuit with tensor network contraction mode using JDtensorPath.

\begin{lstlisting}[caption={How to use tensor network contraction mode.}, captionpos=b, label={tn_mode}, language={Python}]
def circuitDef(parameters):
    # Circuit definition
# parse and construct the circuit
circuit = tedq.circuit(circuitDef, n_qubits, parameters)
# compile the quantum with tensor network contraction mode by using JDtensorPath
from jdtensorpath import JDOptTN as jdopttn
slicing_opts = {'target_size':2**28, 
                'repeats':512, 
                'target_num_slices':None, 
                'contract_parallel':True}
                
hyper_opts = {'max_time':120, 
             'max_repeats':128, 
             'search_parallel':8, 
             'slicing_opts':slicing_opts}
             
compileCir = circuit.compilecircuit(
                backend = "pytorch", 
                use_jdopttn = jdopttn, 
                hyper_opt = hyper_opts, 
                tn_simplify = False)
\end{lstlisting}

To enable the tensor contraction mode, user needs to specify the options for slicing and hypergraph-based optimizer, shown as ``$slicing\_opts$" and ``$hyper\_opts$", and pass these options and a contraction path optimizer, shown as ``$jdopttn$", into the circuit compiler. The compiled circuit would be constructed in tensor contraction mode. The detailed description of each option and argument can be found in the github repository and the documentation of TeD-Q.


\section{Variational quantum circuit}\label{section_vqc} 
 Optimization is an important common factor of most NISQ algorithms, including quantum machine learning and quantum chemistry simulation. It relies on a hybrid quantum-classical optimization process on the parameterized quantum circuit, or the so-called variational quantum circuit (VQC). For instance, quantum neural network (QNN) can be regarded as a VQC. In quantum machine learning tasks, a VQC could work as a component together with some pre- and post-processing like feature extraction and prediction. Typically, the cost function of a task can be taken as the expectation value of a VQC:
 
 \begin{equation}
     f(x;\theta) = \langle\hat{O}\rangle = \langle0|U^{\dag}(x;\theta)\hat{O}U(x;\theta)|0\rangle.
 \end{equation}
 The optimization process relies on the gradient of the circuit output with respect to the parameters of the quantum gates. The parameter shift and finite differential methods are the common methods to obtain gradients of a quantum circuit for both quantum hardware and simulation, which are the built-in methods of TeD-Q. However, the simulation with such methods usually takes a very long time due to the computational complexity. Meanwhile, the demand for fast simulation of circuits with large number of qubits has become urgent due to recent development of large-scale quantum devices. To this end, TeD-Q provides an efficient method to compute the gradient of a circuit in the simulation via the conventional backpropagation technique and the newly-developed tensor-network engine of TeD-Q.

\subsection{Backpropagation method} 
The backpropagation method in TeD-Q is based on the well-developed automatic differentiation libraries, such as PyTorch\cite{paszke2017automatic} and JAX\cite{bradbury2021jax}, which have been widely used in classical machine learning. These libraries are integrated into TeD-Q as simulation backends. When the backend is specified, TeD-Q will convert the matrix data of the quantum circuit to the corresponding format. While evaluating the quantum circuit, known as the "forward" computation, the backend will store all of the intermediate results. The backend will use the chain rule to obtain the gradient of the gate parameters from the circuit output, known as backpropagation. This function will store the gradient of each intermediate result so that the gradient of all of the parameters can be evaluated at one time. Although the backpropagation method cannot be applied to real quantum devices, this is the fastest method to obtain the gradient in quantum simulations. Thus, this is the default method for the simulation backends.  

\begin{lstlisting}[caption={How to use backpropagation method to get the gradient of parameters.}, captionpos=b, label={how_to_use_back_prop}, language={Python}]
def circuitDef(parameters):
    # Circuit definition
# parse and compile the circuit
cir = tedq.circuit(circuitDef, n_qubits, parameters)
myCompileCir = cir.compilecircuit(
                    backend="pytorch")
\end{lstlisting}

\subsection{Parameter shift and finite differential method} 
In real quantum devices, the gradients of the variables can only be obtained by the parameter shift method or finite differential method. The parameter shift method can provide the analytical derivatives, while the finite-differential method provides only numerical derivatives. 

In TeD-Q, we implement the parameter shift method based on the theoretical paper ~\cite{param_shift,four_term_parameter_shift} and recent developments by \cite{bergholm2018pennylane}. This method can be applied to quantum gates with the form $e^{i\mu H}$, where $H$ is a Hermitian operator with two or four eigenvalues, and $\mu$ is the parameter of the gate. For a single gate parameter, the analytical derivative can be calculated by evaluating the expectation value of the circuit twice with the parameter value being $\mu +s$ and $\mu -s$\cite{param_shift}, as shown in Equation~\ref{eq_finite_diff_n_param_shift}, where $s$ is the shift and $k$ is the shift constant. TeD-Q's backend will shift the parameters automatically with proper spacing for each gate. For the simulation backend, like PyTorch or JAX, the user can use the parameter shift method to obtain the gradient by specifying it in the differential option while compiling the circuit, as shown in Code Listing~\ref{params_shift}. For the hardware backend, the parameter shift is the default and only method to obtain the gradient.

\begin{equation}
\label{eq_finite_diff_n_param_shift}
\frac{\partial f}{\partial \mu}=\left\{
    \begin{array}{lr}
    k[f(\mu+s)-f(\mu-s)] & \text{parameter \ shift} \\
    \\
    \frac{f(\mu+\Delta\mu)-f(\mu)}{\Delta\mu} & \text{finite \ differential}
    \end{array}
    \right.
\end{equation}

\begin{lstlisting}[caption={How to use parameter shift method to get the gradient of parameters.}, captionpos=b, label={params_shift}, language={Python}]
def circuitDef(parameters):
    # Circuit definition
# parse and compile the circuit
cir = tedq.circuit(circuitDef, n_qubits, parameters)
compileCir = cir.compilecircuit(
                    backend="pytorch", 
                    diff_method = "param_shift")
\end{lstlisting}

\subsection{Quantum to classical computing interface} 
The design paradigm of TeD-Q provides seamless integration of quantum and classical processing. In this hybrid algorithm, a quantum circuit that is parameterized by tunable variables is seen as a normal Python function with a specific interface. It can be inserted into a classical optimization function or wrapped in a layer of a machine learning framework module.

TeD-Q supports two interfaces -- JAX array and PyTorch tensor. For the JAX simulation backend, the JAX array is the default one, but the PyTorch tensor interface can also be used via the interface keyword argument (interface="pytorch") while compiling the quantum circuit. As for the PyTorch simulation backend and hardware backend, TeD-Q only supports the PyTorch backend at the time of writing. The interfaces provide a seamless connection to the classical part in "forward" calculation and in "backward" gradient extraction. For simulation backends with their own interface and default backpropagation gradient method, the integration is straightforward, relying on built-in data structure and automatic differentiation algorithm of JAX or PyTorch. For other cases (JAX backend with PyTorch tensor interface, hardware backend, parameter shift method), the classical backpropagation algorithm cannot process the quantum information inside the quantum circuit. For this situation, the quantum circuit is treated as a black box. TeD-Q implements custom autograd functions and will compute and provide the gradient (one output) or Jacobian (multiple inputs and multiple outputs) of the quantum circuit with respect to its classical inputs and variables. Code Listing~\ref{interface} shows an example of connecting the JAX backend quantum circuit to the classical calculation by the PyTorch tensor interface. More detailed examples can refer to Section~\ref{section_hardware_use_case}. The integration is imperceptible to the user, via JAX or PyTorch, with the flexibility to run the hybrid quantum-classical algorithms code on CPUs, GPUs, and TPUs.

\begin{lstlisting}[caption={Quantum to classical computing interface. A quantum circuit is compiled into an executable quantum object by using JAX simulation backend and PyTorch tensor interface. The executable quantum circuit is then integrated with classical cost function.}, captionpos=b, label={interface}, language={Python}]
# Quantum circuit construction
    # Neglect here

# Compile quantum circuit with JAX quantum simulation backend and PyTorch interface
compileCirJAX = circuit.compilecircuit(
                        backend="jax",
                        interface="pytorch")

# Connect quantum systems to classical cost function by PyTorch interface
def cost(weight, params):
    return weight * compileCirJAX(params)

# PyTorch tensor is used as input
weight = torch.tensor(0.5)
params = torch.tensor([0.12])
cost(weight, params)
\end{lstlisting}

\section{Use case}
The detailed introduction and manual of the user interface can be found in the online documentation. It includes step-by-step examples demonstrating the feature of Ted-Q. In the following section, We briefly go through the main features and its use cases.

\subsection{Basic quantum circuits} 
Quantum circuit evaluation is the basic function of TeD-Q. It provides a universal set of quantum gates so all kinds of quantum circuits can be constructed and simulated. With the tensor network engine described in Section~\ref{section_tnc}, TeD-Q can process the simulation faster than the conventional quantum simulators. Besides, TeD-Q is equipped with an easy-to-use graphical circuit composer and circuit drawer. These tools are well integrated with TeD-Q's simulator. Users can use it for quick prototyping of the quantum circuit. Moreover, the hardware backend in Ted-Q provides an interface to the real quantum devices like Qiskit\cite{qiskit}. Users Can switch the simulation to a real experiment with a simple command. The following sections demonstrate these features via TeD-Q API.

\subsubsection{Fast simulation} 
Consider a simple two-qubit quantum circuit with a $Hadamard$ gate on qubit $0$, a $CNOT$ gate on qubits $0$ and $1$, and a rotation gate $Rot$ on qubit $1$ apply to the initial zero state. And we want to obtain its quantum state with some input rotation angles. TeD-Q can implement the simulation as code block~\ref{fast_simulation}. Qubits are counted from 0 and multiqubit registers labeled with the first (zeroth) qubit on the left. Measurement of a quantum state is only available by the simulation backend, while both simulation and hardware backend enables expectation values and probability measurements. Keyword argument ``requires\_grad" can disable the gradient feature and improve the simulation speed since the backend doesn't need to store the intermediate state for backpropagation.

\begin{lstlisting}[caption={An example of a simple two-qubits quantum circuit. $Hadamard$ and $CNOT$ gates are used to generate a entangled state, $Rot$ gate is used for adding some rotation angle and the final quantum state is measured at the end.}, captionpos=b, label={fast_simulation}, language={Python}]
import tedq
import torch
n_qubits=2

# Circuit definition
def circuitDef(phi, theta, omega):
    tedq.Hadamard(qubits=[0])
    tedq.CNOT(qubits=[0, 1])
    tedq.Rot(phi, theta, omega, qubits=[1])
    # Return quantum state
    return tedq.measurement.state()

phi = torch.tensor([0.4])
theta = torch.tensor([0.5])
omega = torch.tensor([0.6])
circuit = tedq.Circuit(circuitDef, n_qubits, phi, theta, omega)
# Parse and compile the circuit 
compileCir = circuit.compilecircuit(backend="pytorch", requires_grad=False)

# Obtain the quantum state
quantum_state = compileCir()
\end{lstlisting}

\subsubsection{Intuitive prototyping}
\begin{figure}[!h]
    \centering
    \includegraphics[width=0.45\textwidth]{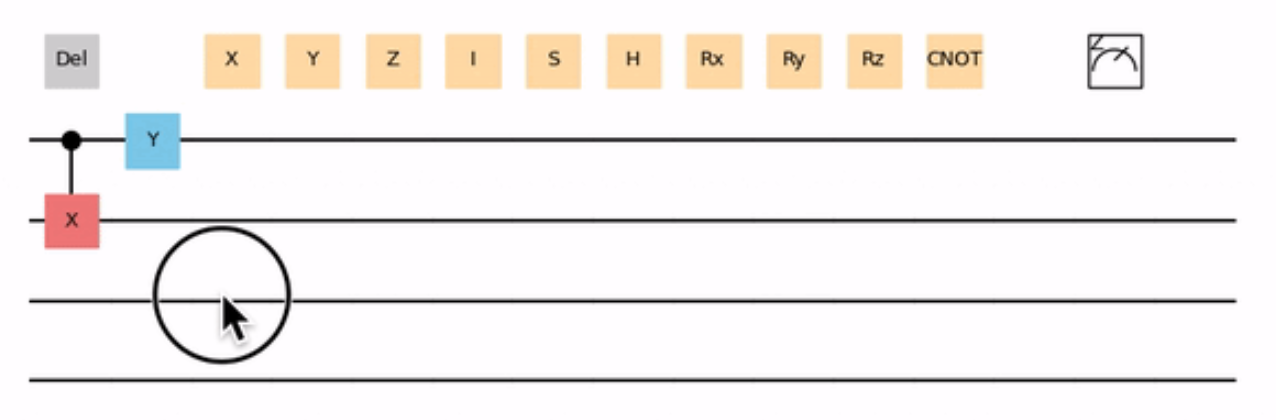}
    \caption{Circuit composer} 
    \label{circuit_composer}
\end{figure}

Circuit composer in TeD-Q is a WYSIWYG (what you see is what you get) editor for a quantum circuit. The graphical composer can be initialized with a single line command, as shown in Code Listing~\ref{init_composer}. It consists of a panel for the template quantum gates and the several qubits indicated by lines, as shown in Figure~\ref{circuit_composer}. Users can construct the circuit by dragging the icon of the quantum gate to a specific qubit line or remove one quantum gate by dragging it away from the qubit line. Users can also adjust the sequence of the operations by pulling the icons to the desired location. 

\begin{lstlisting}[caption={Usage of circuit composer}, captionpos=b, label={init_composer}, language={Python}]
# Enable circuit composer
composer = tedq.circuit_composer(4, figsize=(11,5))
\end{lstlisting}

The composing result can be converted and fed into TeD-Q's tensor network engine. The user can obtain the simulation result using the same method mentioned in the previous section, as shown in the Code Listing~\ref{composer_to_circuit}. Users can also get the circuit definition from the composing result for further development with TeD-Q API.

\begin{lstlisting}[caption={Converting the composing result to a TeD-Q circuit and obtaining the circuit definition}, captionpos=b, label={composer_to_circuit}, language={Python}]
# Convert to circuit class
circuit = composer.toCircuit()
compiledCircuit = circuit.compilecircuit()

# Evaluating the circuit
prob = compiledCircuit()

# Print the circuit definition
print(circuit)
\end{lstlisting}

Besides the circuit composer, the circuit initialized by the script can also be visualized with the circuit drawer, as shown in the Code Listing~\ref{circuit_drawer}. It follows the same drawing style as the circuit composer. Moreover, the function will automatically fit the circuit to the correct size of the plot.
\begin{lstlisting}[caption={Usage of circuit drawer}, captionpos=b, label={circuit_drawer}, language={Python}]
# Init a circuit
circuit = tedq.circuit(circuitDef())
# Draw a circuit
tedq.draw_circuit(circuit)

\end{lstlisting}

\subsubsection{Hardware compatibility} \label{section_hardware_use_case} 
Aside from simulation, the circuit can also be adopted to real quantum devices with the hardware backend. At the time of writing, TeD-Q only supports IBM-Q quantum devices, and more quantum hardware devices will be available soon. The built-in parameter shift method is responsible for obtaining gradients of the quantum circuit, and the PyTorch tensor interface is provided for communication with other functions. Code Listing~\ref{hardwarebackend} shows an example of compiling a quantum circuit into two executable quantum objects with both simulation and hardware backends and tries to minimize their difference by tuning input parameters with a built-in gradient-descent optimizer.

\begin{lstlisting}[caption={Hardware compatibility example. A quantum circuit is compiled into two executable quantum objects with JAX simulation backend and hardware backend. And these two executable quantum circuit are connected to classical cost function with PyTorch interface.}, captionpos=b, label={hardwarebackend}, language={Python}]
# Quantum circuit construction
    # Neglect here

# Compile quantum circuit with hardware backend, default PyTorch interface
from qiskit import IBMQ
IBMQ.enable_account('your IBMQ token')
compileCirHardware = circuit.compilecircuit(backend="IBMQ_hardware")

# Compile quantum circuit with JAX quantum simulation backend and PyTorch interface
compileCirJAX = circuit.compilecircuit(backend="jax", interface="pytorch")

# Connect two quantum systems to classical cost function by PyTorch interface
def cost(params):
    results = (compileCirHardware(params) - compileCirJAX(params))**2
    return results
    
# Tuning variables with built-in optimizer
Optimizer = tedq.GradientDescentOptimizer(
                objective_fn=cost, 
                trainable_params=[0, 1], 
                interface="pytorch")
params = (torch.tensor([0.12], requires_grad=True), 
          torch.tensor([0.75], requires_grad=True))
new_variables = Optimizer.step(*params)
\end{lstlisting}

\subsection{Variational quantum algorithm} 
A variation quantum algorithm consists of a variational quantum circuit and a classical optimizer. The method to obtain the gradient of a gate parameter in a quantum circuit is described in Section~\ref{section_vqc}. Recursively updating the gate parameters allows the circuit to be optimized according to a certain cost function. It's widely used in current NISQ devices and quantum simulations. With TeD-Q's tensor engine, simulation can be processed at a very high speed compared to other simulation packages. Besides, with the hardware backend, the variational algorithm can also be adapted to a real quantum device. The following example shows a simple circuit based on the variational quantum algorithm to demonstrate how to construct a VQA circuit by TeD-Q API.

\begin{lstlisting}[caption={Example of variational quantum algorithm.}, captionpos=b, label={vqe_example}, language={Python}]
import tedq
import torch
n_qubits=3

# Circuit definition
def circuitDef(theta):
    # Circuit
    tedq.RX(theta[0], qubits=[0])
    tedq.RY(theta[1], qubits=[1])
    tedq.RZ(theta[2], qubits=[2])
    
    tedq.CNOT(qubits=[0, 1])
    tedq.CNOT(qubits=[1, 2])
    tedq.CNOT(qubits=[2, 0])
    
    # Measurement
    for idx in range(n_qubits):
        tedq.measurement.expval(tedq.PauliZ(qubits=[idx]))

# Parse and compile the circuit        
circuit = tedq.Circuit(circuitDef, n_qubits, torch.rand(n_qubits))
my_compilecircuit = circuit.compilecircuit(backend="pytorch" )

# Init the parameters by the torch library
params = torch.rand(n_qubits, requires_grad=True)

# Forward computation of the expectation value
exp_val = my_compilecircuit(params)

# Define the loss function
optimizer = torch.optim.Adam([params], lr=0.5)
loss = torch.nn.MSELoss(reduction='mean')
l = loss(exp_val, torch.Tensor([0, 1, 1]))

# Backward computation for the gradient of the parameters
l.backward()

# Output the result
print("Expected value from circuit: ", exp_val)
print("Gradient of parameters", params.grad)
\end{lstlisting}

\subsection{Quantum machine learning} 
Machine learning deals with the problem of generalization from finite data in high-dimensional Hilbert spaces. It is a very appealing application for quantum computation since both of them involve statistics theories at a fundamental level. Users can transform classical feature information into the quantum state through encoding schemes (basis encoding, amplitude encoding, angle encoding, etc.):
\begin{equation}
|\psi(x)\rangle=\left\{
    \begin{array}{lr}
    |i_x\rangle & \text{basis \ encoding} \\
    \\
    \sum_i^n x_i|i\rangle & \text{amplitude \ encoding} \\
    \\
    \otimes_i^n R(x_i)|0^n\rangle & \text{angle \ encoding}
    \end{array}
    \right.
\end{equation}
VQA can be designed as a linear model to classify data explicitly in Hilbert space. With a nice performance of TeD-Q's tensor engine, a more complex VQA circuit can be constructed and simulated, which allows testing of a large-scale quantum and hybrid quantum-classical machine learning algorithm with the simulator. 
\subsubsection{Templates} 
TeD-Q equips several pre-defined quantum circuit templates, or ansatz, commonly used in the quantum machine learning field to enable the fast development of the circuit for quantum machine learning. The following briefly introduces three templates in TeD-Q -- Random-generated layer, Fully-connected layer, and Hardware-efficient layer.

\begin{figure}[!h]
    \centering
    \includegraphics[height=0.125\textheight]{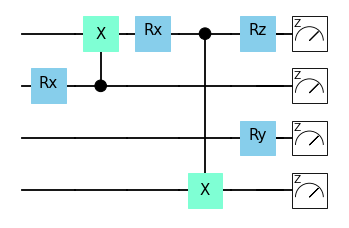}
    \caption{Randomly-generated circuit} 
    \label{randomly_generated}
\end{figure}

The randomly-generated layer can provide a circuit with a single- and two-qubits gate while their locations and parameters are random (Fig.~\ref{randomly_generated}). It's suitable for simulating a non-linear transformation, as shown in TeD-Q's Quanvolution Neural Network example.

\begin{figure}[!h]
    \centering
    \includegraphics[height=0.114\textheight]{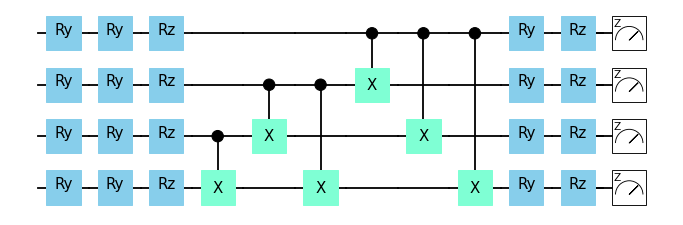}
    \caption{Fully-connected circuit} 
    \label{fully_connected}
\end{figure}

The fully-connected layer is a parameterized circuit to provide maximum entanglement among each qubit. The number of layers is called depth, which the user can customize. Each qubit is connected to the other qubits in each layer by a CNOT gate followed by a two-parameterized rotation gate (Fig.~\ref{fully_connected}). This template is usually used to simulate a classical fully-connected layer in quantum machine learning. 

\begin{figure}[!h]
    \centering
    \includegraphics[height=0.125\textheight]{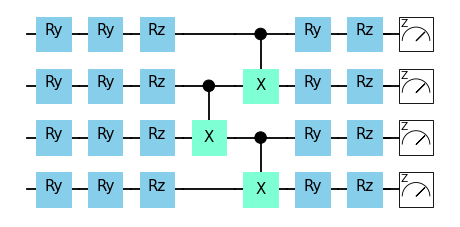}
    \caption{Hardware efficient circuit} 
    \label{hardware_efficient}
\end{figure}

The hardware-efficient layer is a practical implementation of a fully-connected circuit. In the real quantum device, a two-qubit operation is only efficient between the qubit nearby the qubit. Therefore, the CNOT gates in this template are only placed between the qubits nearby a selected qubit (Fig.~\ref{hardware_efficient}).

\subsubsection{Quantum layer wrapping}
Users can easily exploit quantum machine learning by replacing some layers of a well-developed classical machine learning model with a wrapped quantum layer. Code Listing~\ref{quantum_layer} illustrates how to build the pipeline in TeD-Q.
\begin{lstlisting}[caption={Code snippet for creating a hybrid-quantum-classical machine learning model for image recognition. The quantum circuit is wrapped into a quantum layer and used to replace the fully-connected layer of the classical deep learning model ResNet-50. The main structure of this example follows the PennyLane tutorial on quantum transfer learning\cite{bergholm2018pennylane}}, captionpos=b, label={quantum_layer}, language={Python}]
# Global variable
n_qubits = 8      # Number of qubits
q_depth = 12      # Number of variational layers
n_classes = 10    # Number of image classes

# Angle encoding
def features_encoding(nqubits, w):
    for idx in range(nqubits):
        tedq.Hadamard(qubits=[idx])
    for idx, element in enumerate(w):
        tedq.RX(element, qubits=[idx])

# Define a variational quantum circuit
def circuitDef(q_input_features, q_weights_flat):
    # Reshape weights
    q_weights = q_weights_flat.reshape((q_depth+1)*2, n_qubits)

    # Encode classical features into quantum state
    features_encoding(n_qubits, q_input_features)

    # Built-in template circuit
    tedq.templates.HardwareEfficient(n_qubits, q_depth, q_weights)
    
    # Expectation values measurement
    exp_vals = [tedq.expval(tedq.PauliZ(qubits=[position])) for position in range(n_qubits)]
    return tuple(exp_vals)
            
# Quantum circuit construction
circuit = tedq.Circuit(circuitDef, n_qubits, torch.zeros(n_qubits), torch.zeros((q_depth+1)*2, n_qubits))
compiledCircuit = circuit.compilecircuit(backend="pytorch")

# Quantum layer with PyTorch NN interface
class DressedQuantumNet(nn.Module):
    def __init__(self):
        super().__init__()
        self.pre_net = nn.Linear(2048, n_qubits)
        self.q_params = nn.Parameter(q_delta * torch.randn((q_depth+1) * n_qubits*2))     
        self.post_net = nn.Linear(n_qubits, n_classes)
        def q_func(q_input_features):
            return compiledCircuit(q_input_features, self.q_params)
        from functorch import vmap
        # for batch executation
        self.batched_q_func = vmap(q_func)

    # Tensors moves through quantum circuit
    def forward(self, input_features):
        pre_out = self.pre_net(input_features)
        q_in = torch.tanh(pre_out) * np.pi / 2.0
        q_out = self.batched_q_func(q_in).float()
        return self.post_net(q_out)

# Replacing classical machine learning model ResNet-50 fully-connected layer with quantum layer     
model_hybrid = torchvision.models.resnet50()
model_hybrid.fc = DressedQuantumNet()
\end{lstlisting}

The above example shows a way of constructing a hybrid quantum-classical machine learning model. TeD-Q can straightforwardly integrate quantum algorithms with various well-established machine learning practices. The quantum circuit model behaves as if it was a pure classical one. All the following processes, like cost function or optimization, can be seamlessly adopted. In the next , a pure quantum machine learning example is introduced and used as a benchmark to compare the performance of TeD-Q and other software.

\section{Performance} 
\label{Performance}
We benchmark TeD-Q's performance with other quantum software through two examples -- multiple qubits rotation (MQR) and many body localization (MBL)~\cite{nandkishore2015many}. MQR is a very simple algorithm which uses hardware-efficient network to flip the quantum state from all zeros state $|0\rangle^{\otimes n}$ to all ones state $|1\rangle^{\otimes n}$. It is suitable for evaluating TeD-Q's performance on small and simple circuit which can be run on personal computer.
MBL is a relatively complex problem of great significance both theoretically and practically. It is used to demonstrate the great advantage of TeD-Q tensor network contraction mode. The source code of the comparison can be found on TeD-Q's Github.

\subsection{Multiple qubits rotation} 
\begin{figure}[!h]
    \centering
    \includegraphics[width=0.5\textwidth]{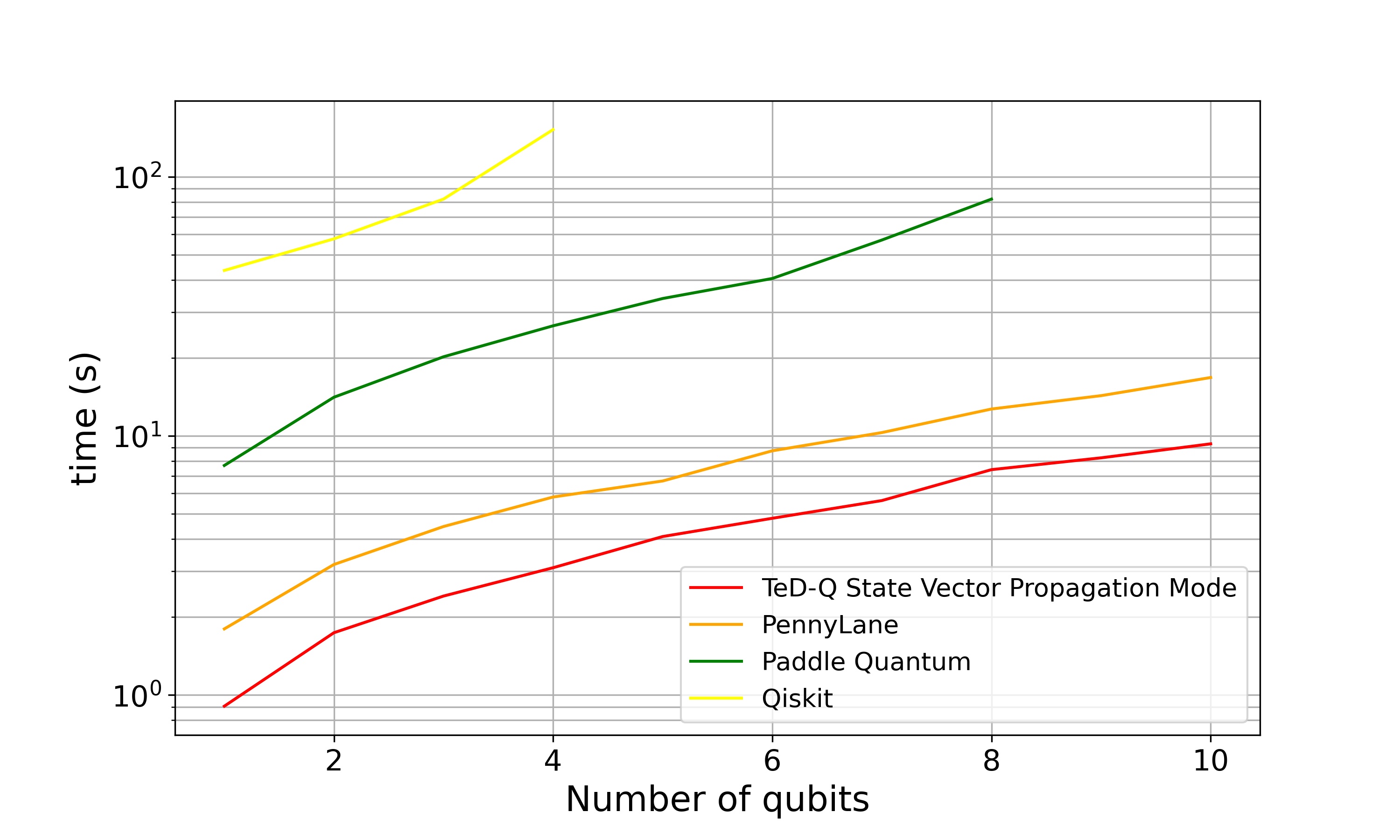}
    \caption{Performance comparison among TeD-Q, PennyLane, Paddle Quantum and Qiskit for small simple circuit on single laptop CPU. TeD-Q result uses the PyTorch backend with backpropagation gradient method.} 
    \label{CPU_small}
\end{figure}

In this simulation, the initial quantum state is set to be all zeros state $|0\rangle^{\otimes n}$ followed by a variational hardware-efficient network and expectation value measurement on Pauli-Z basis for each qubit. The learning is done by minimizing sum of the measurements through tuning parameters via gradient descent method. And the performance of platforms are evaluated by the time cost for finishing 500 iterations of training. Since there is a great demand for user to develop and learn quantum algorithm through simple circuit on their personal computer, the evaluation is done with single AMD Ryzen$^{TM}$ 5 3500U mobile processor CPU. Fig.~\ref{CPU_small} shows the advantage of TeD-Q in this application.

\subsection{VQE of Hydrogen Molecule}
Variational quantum eigensolver (VQE) method is one of the popular applications of quantum computing in the NISQ era~\cite{preskill2018quantum}. It is a concrete example of variational quantum algorithm (VQA) used for solving quantum chemistry problems, specifically, the eigenstate problem. In a VQE algorithm, a parameterized quantum circuit would prepare a quantum state, dubbed as the ansatz state, which is supposed to be the ground state of the Hamiltonian of the molecule. By repeatedly measuring the ansatz state with different measurement settings, the expectation value of the Hamiltonian could be evaluated. An optimization process could be applied to the parameters of the quantum circuit so that the expectation value could be minimized. 

Here we use the VQE of Hydrogen molecule to illustrate the performance of TeD-Q versus other platforms. In this experiment, there are totally four qubits and the quantum circuit contains 12 freely changing parameters. (The problem can in principle be simplified to use just one qubit, but we did not do this since we only want to compare the speed on the same ground.) 

\begin{figure}[!h]
    \centering
    \includegraphics[width=0.45\textwidth]{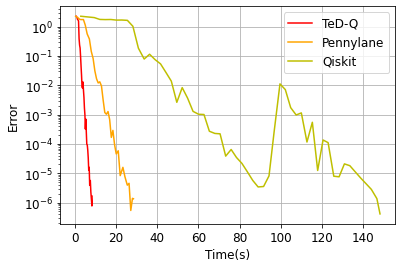}
    \caption{Error between the VQE result and the FCI reference energy.} 
    \label{fig_vqe_speed_comparison}
\end{figure}

\begin{figure}
     \centering
     \begin{subfigure}[b]{0.4\textwidth}
         \centering
         \includegraphics[width=\textwidth]{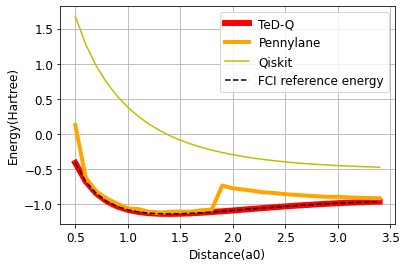}
         \caption{Execution time limited to 5 seconds.}
         \label{fig_vqe_energy_distance_5s}
     \end{subfigure}
     \hfill
     \hfill
     \begin{subfigure}[b]{0.4\textwidth}
         \centering
         \includegraphics[width=\textwidth]{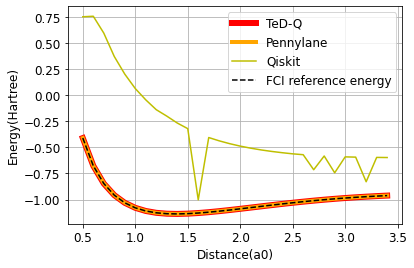}
         \caption{Execution time limited to 30 seconds.}
         \label{fig_vqe_energy_distance_30s}
     \end{subfigure}
     \caption{Energy curve of the Hydrogen atom from VQE with limited execution time and FCI methods.}
\end{figure}

We run the experiment on a Macbook Air with Intel Core$^{TM}$ i5-1030NG7 CPU. As we could see from Fig.~\ref{fig_vqe_speed_comparison}, TeD-Q converges to an error of $10^{-6}$ Hartree about 3 times faster than Pennylane and around 15 times faster than Qiskit. With a limited time of execution, for instance, within 5 seconds as shown in Fig.~\ref{fig_vqe_energy_distance_5s}, TeD-Q can correctly plot the energy curve of the ground state with different H-O distances, while Pennylane can almost do it and Qiskit being far away from that. When the time limit is set to 30 seconds Pennylane can converge to the correct energy curve but Qiskit still cannot, as shown in Fig.~\ref{fig_vqe_energy_distance_30s}. From this comparison, we can see TeD-Q indeed has advantages in simulating traditional VQE algorithm over the compared platforms.

\subsection{Quantum GAN}
Classical generative adversarial network (GAN) has found its position in artificial intelligence via rapid growing applications recent years~\cite{goodfellow2020generative}. The idea is to utilize a generative network to create new data that resembles the original data, while a discriminative network tries to evaluate the distance. The two networks contest with each other to play a zero-sum game, which leads to improved quality of the newly generated data.

Quantum GAN borrowed this idea and wishes to boost the performance by quantum computing~\cite{tian2022recent,huang2021experimental}. While it is possible to use QNN in both the generative and discriminative networks, here, we adopt both the patch and non-patch quantum GAN methods where only the generative part is made quantum. 

\begin{figure}
     \centering
     \begin{subfigure}[b]{0.45\textwidth}
         \centering
         \includegraphics[width=\textwidth]{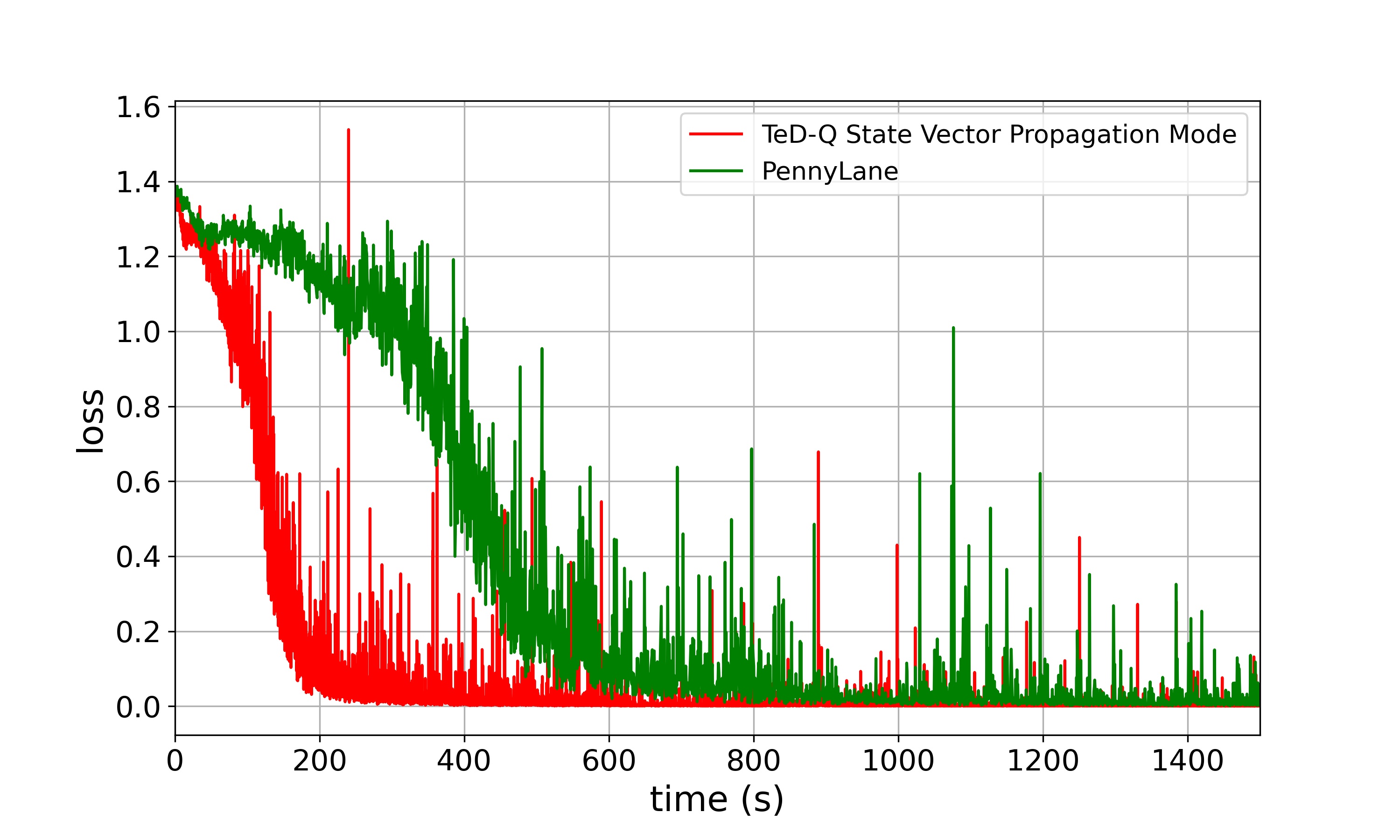}
         \caption{QGAN patch}
         \label{fig:y equals x}
     \end{subfigure}
     \hfill
     \begin{subfigure}[b]{0.45\textwidth}
         \centering
         \includegraphics[width=\textwidth]{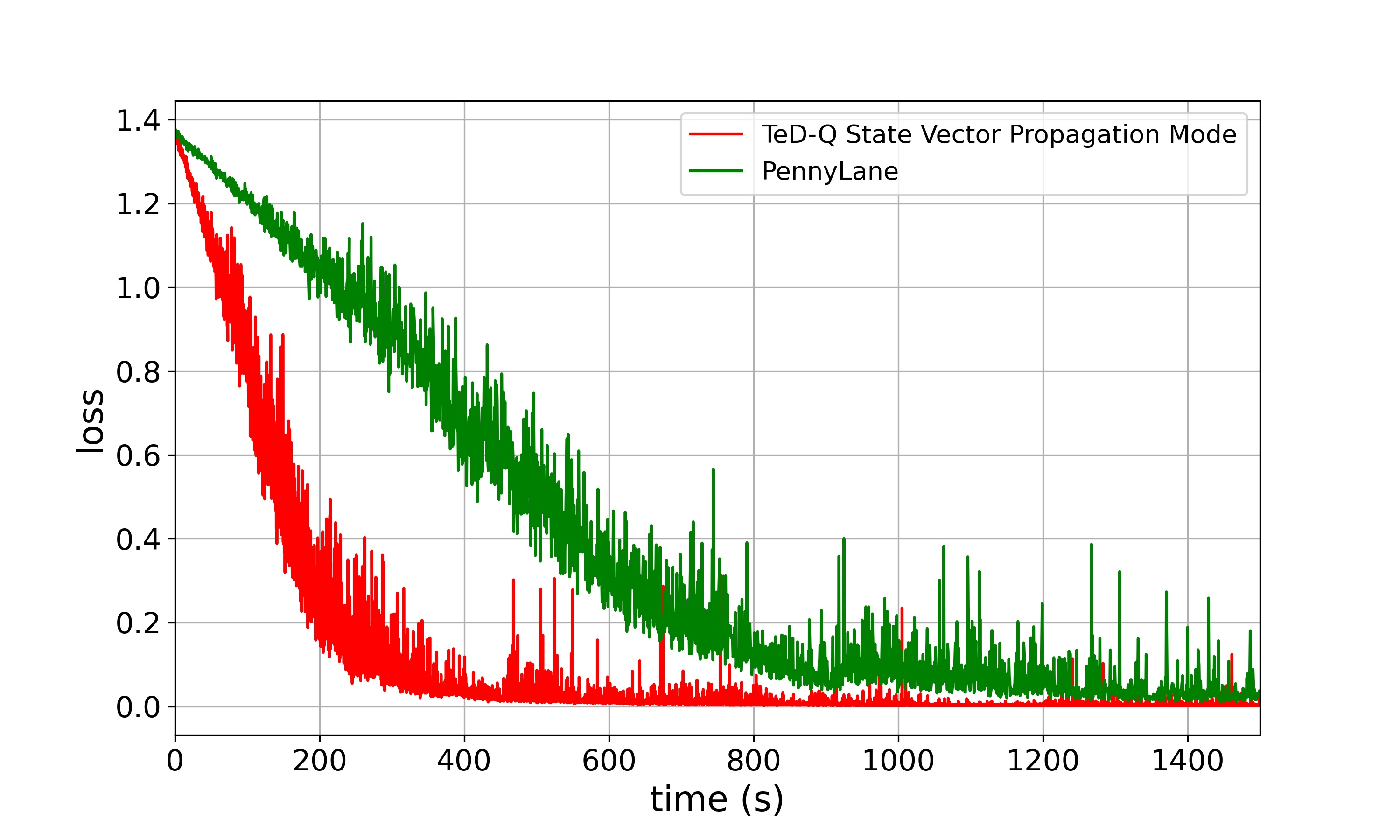}
         \caption{QGAN non-patch}
         \label{fig:three sin x}
     \end{subfigure}
     
        \caption{Performance comparison between TeD-Q and PennyLane for quantum GAN training on single laptop CPU. TeD-Q result uses the PyTorch backend with backpropagation gradient method.}
        \label{qgan}
\end{figure}

In this example, the generator tries to mimic the handwritten zeros from a small set of grayscale samples images with size $8\times 8$. Please see~\cite{huang2021experimental} for details of the QGAN example. Similar to the multiple qubits rotation example, we run the experiment with an AMD Ryzen$^{TM}$ 5 3500U mobile processor CPU. As we can see from the result Fig.~\ref{qgan}, in either patch or non-patch scenario, TeD-Q outperforms Pennylane in the training of the quantum GAN model.

\subsection{Quantum  Architecture Search}
Quantum architecture search (QAS) was recently proposed to generate better variational ansatz circuits for given VQA tasks~\cite{du2022quantum,linghu2022quantum}. By constructing an ansatz pool, QAS iteratively samples a new ansatz at every step of the optimization in the VQA. It can be shown that by doing this, the resulting ansatz improves the performance of the VQA in terms of both circuit output and learnability in learning tasks.

\begin{figure}[!h]
    \centering
    \includegraphics[width=0.4\textwidth]{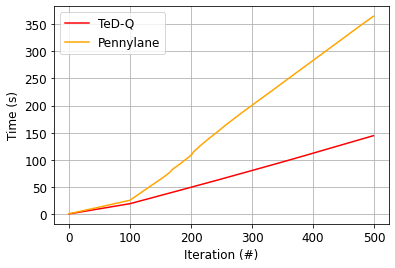}
    \caption{QAS performance comparison between TeD-Q and Pennylane} 
    \label{fig_qas}
\end{figure}

In the training of the QAS task, the ansatz circuit is replaced by new sample from the pool at every step of the optimization, which requires the backend to compile the new ansatz circuit at every step. Thus, the difference of the performance mainly depends on the speed of implementing each circuit. We run the experiment on a Macbook Air with Intel Core$^{TM}$ i5-1030NG7 CPU. The result in Fig.~\ref{fig_qas} shows the time the training takes with increasing iteration numbers. TeD-Q is more than 2 time faster than Pennylane when performing QAS tasks.

\subsection{Many body localization} 
Many body localization (MBL) is a phase of an isolated many-body quantum system that prevents it from thermalization; that is, a system in this phase can preserve the initial information as time evolves. It provides a possibility to hold the quantum effect for a long time. The experimental and theoretical research in this field has been dramatically progressing in recent years due to the rapid development of quantum technology. In this section, we demonstrate that TeD-Q can process a simulation of a 1-D MBL along with a quantum neural network with more than 50 qubits at a reasonable speed, which is not achievable by conventional quantum simulators.

For an isolated many-body quantum system, it evolves under its intrinsic Hamiltonian. The interaction between the particles exchanges information and energy so that the close system reaches thermal equilibrium as the whole system is the thermal bath of its subsystem\cite{mbl_theory}. The information of its initial state is eventually washed out. However, in the presence of sufficient disorder potential, the system can reach the many-body localization phase and prevent it from falling into thermal states even though the particles in the system still interact.

\begin{figure}[!h]
    \centering
    \includegraphics[width=0.4\textwidth]{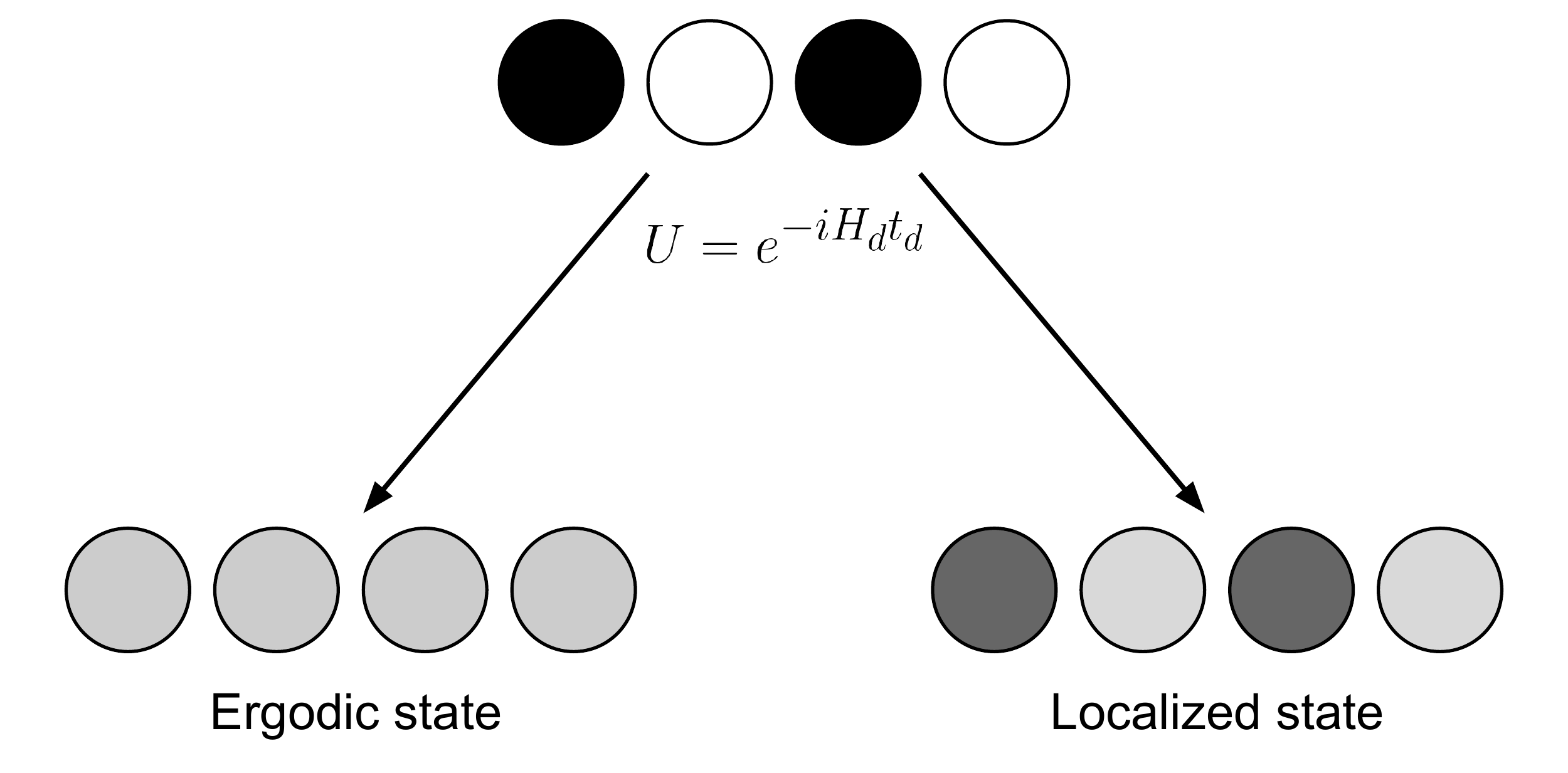}
    \caption{Many body localization. The shade of the circles represents probability to be $|1\rangle$ state. The system is initially prepared in the Neel state. After the evolution under its intrinsic Hamiltonian and a controllable disorder potential, the system can fall into either ergodic state, as shown in the bottom left, or localized state as shown in the bottom right.} 
    \label{fig_mbl_explain}
\end{figure}

The system would fall into thermalization with a small disorder potential, known as the ergodic state. On the other hand, with sufficient disorder power, the system would be in the MBL phase and stay in its initial state, known as the localized state. However, multiple measurements are required to distinguish a localized state from an ergodic state for a many-body system. Following the circuit architecture shown in \cite{mbl2D}, we develop and train a quantum neural network for classifying the states of a 1-D many-body system. With this QNN circuit, we can identify whether the circuit is either an ergodic state or a localized state with a single measurement on the first qubit.

\subsubsection{Simulation setup}
\begin{figure}[!h]
    \centering
    \includegraphics[width=0.5\textwidth]{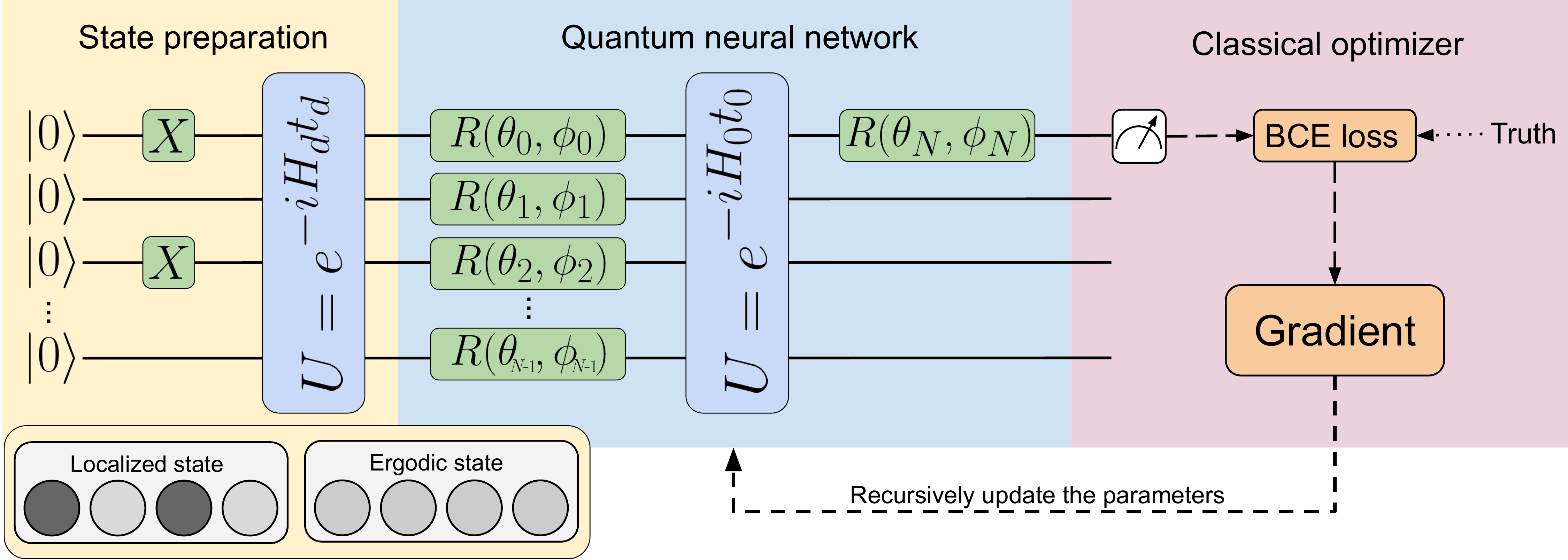}
    \caption{The quantum circuit and the classical optimizer used in MBL simulation. The diagram can be separated into three stages -- state preparation, quantum neural network (QNN), and classical optimizer. The state preparation stage generates the state in either ergodic or localized state. The QNN stage is trained to distinguish ergodic states from localized states. The optimizer stage evaluates the result and optimizes the QNN based on the loss function.} 
    \label{fig_mbl_circuit}
\end{figure}

In our simulation, the qubits are placed in a chain and prepared in the Neel state. It is followed by a simulated Hamiltonian $H_d$ includes the interactions between the nearest qubits and a controllable random disorder $d_i$ for each qubit, which is
\begin{equation}
    H_d = \hbar\sum_{i,j} d_i \sigma_z^i + g_{i,j}(\sigma_x^i \sigma_x^j + \sigma_y^i\sigma_y^j)/2,
\end{equation}
and the system is evolving under this Hamiltonian for a period of time $t_d$. For the state classification, the circuit is then connected to a classifier built in a digital-analog quantum neural network, proposed by \cite{mbl2D}. It is consist of $N_q+1$ single-qubit operation $R(\theta_i, \phi_i)=Z(\theta_i)X(\phi_i)Z(-\theta_i)$, which is the digital part, and an unitary operation $H_0$, which is the analog part, on each qubit.
\begin{equation} 
    H_0 = g_{i,j}(\sigma_x^i \sigma_x^j + \sigma_y^i\sigma_y^j)/2,
\end{equation}
Finally, the only measurement was placed on one of the qubits. The parameters of QNN would be trained so that the measurement of it can classify the state of the system. The circuit diagram is shown in Figure~\ref{fig_mbl_circuit}.

The simulation parameters, like $g_{i,j}$ and $t_d$, follow the setting used in \cite{mbl2D}, which are the properties of a 64-qubit quantum computer. The parameters in the classifier are then trained with the binary cross entropy loss so that the expected value of the the selected qubit can be 1 when the system is in an ergodic state and 0 when the system is in a localized state. 

\subsubsection{Speed comparison}
The performance of TeD-Q is evaluated by the time cost for finishing 50 epochs of 1D-MBL training. The circuit depth is about 89, and each epoch consists of 40 training iterations of the circuit. For the comparison, we also set up the same simulation using Qiskit and Pennylane. CPU simulations use Intel(R) Xeon(R) Processor E5-2680 v4 @ 2.4GHz CPU while GPU simulations use Nvidia Tesla P40 24 GB GPU. 

\begin{figure}
     \centering
     \begin{subfigure}[b]{0.5\textwidth}
         \centering
         \includegraphics[width=\textwidth]{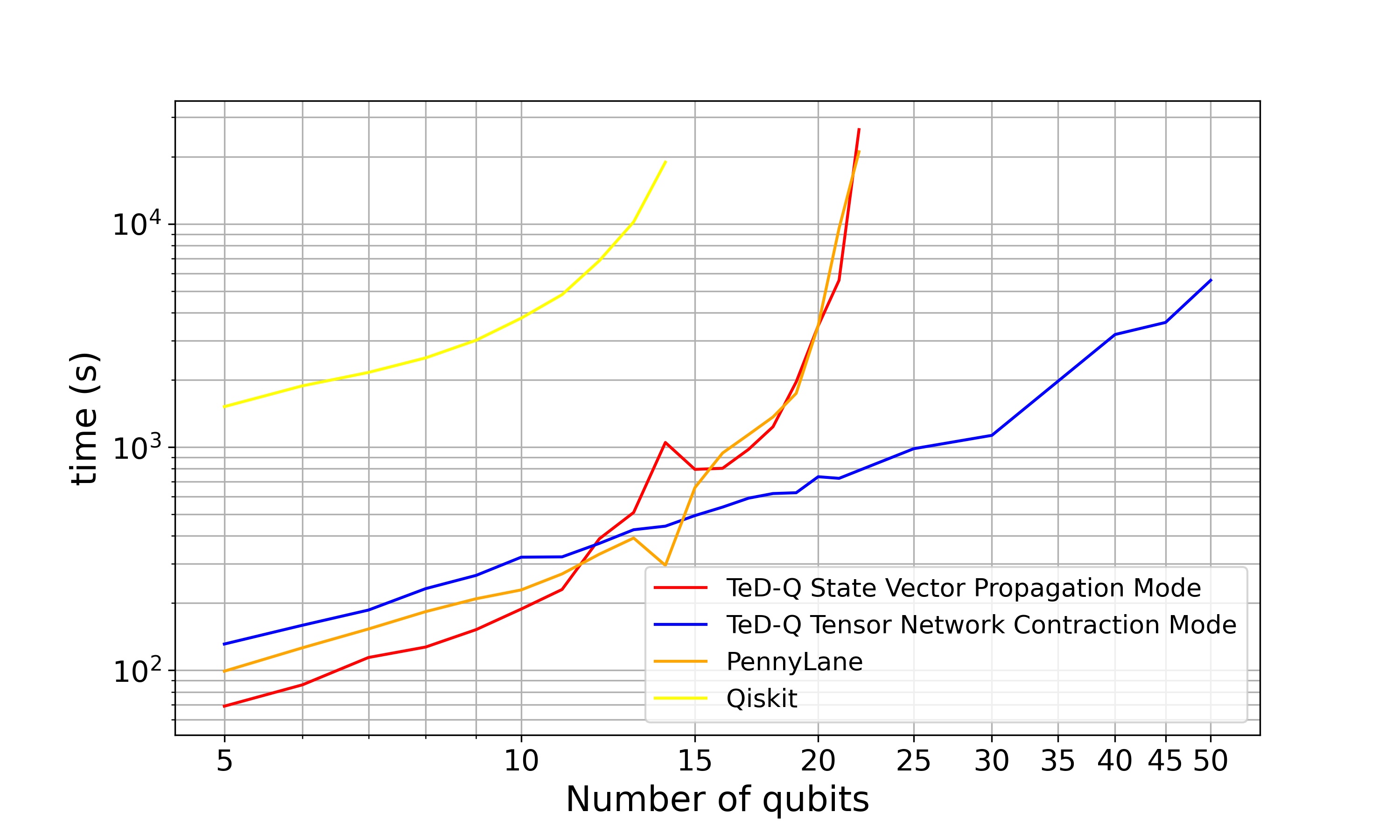}
         \caption{Single CPU mode}
         \label{fig:y equals x}
     \end{subfigure}
     \hfill
     \begin{subfigure}[b]{0.5\textwidth}
         \centering
         \includegraphics[width=\textwidth]{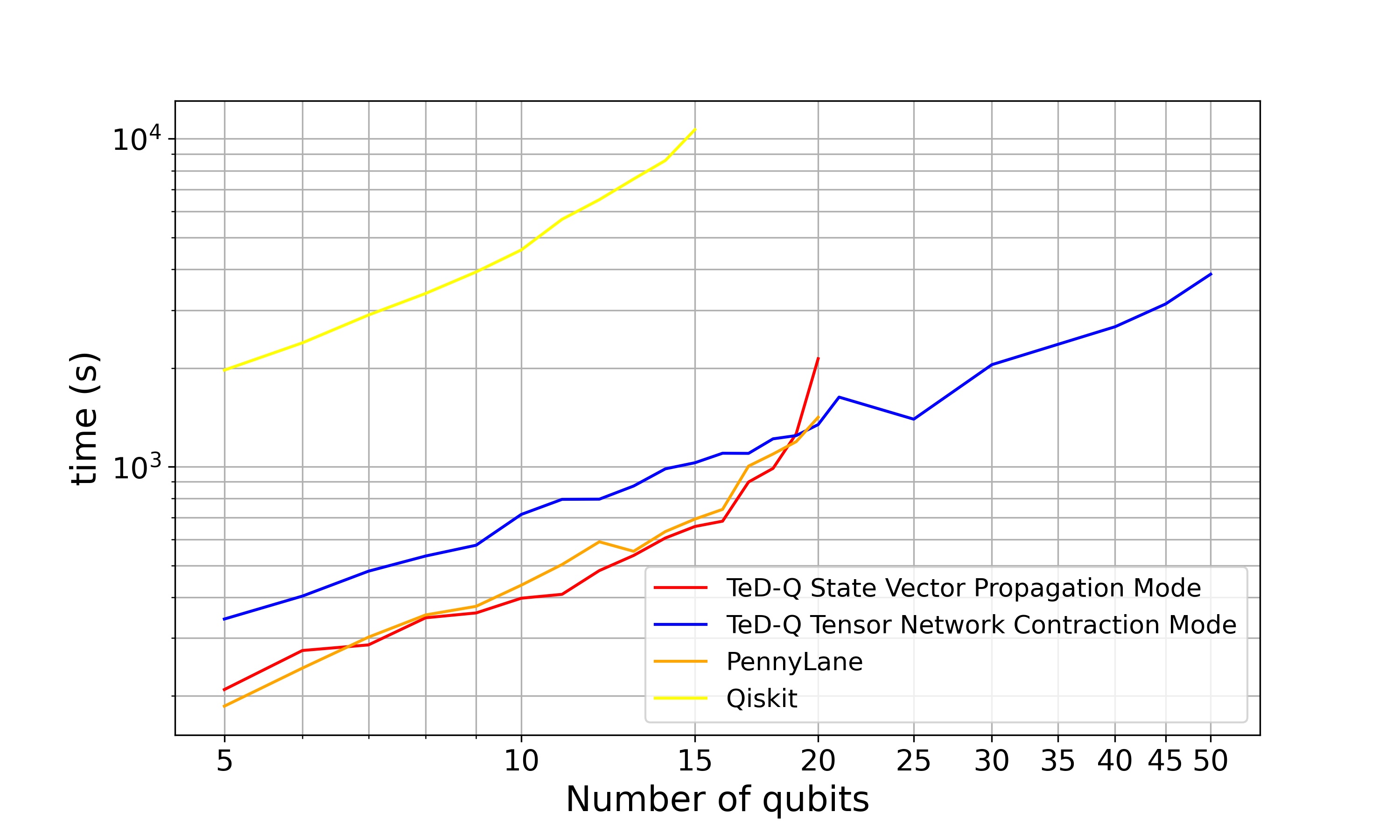}
         \caption{Single GPU mode}
         \label{fig:three sin x}
     \end{subfigure}
     
        \caption{Performance comparison among TeD-Q, PennyLane and Qiskit for the MBL task.}
        \label{performance}
\end{figure}

Figure~\ref{performance} compares these three quantum simulation methods on CPU and GPU. TeD-Q can provide similar behavior as PennyLane while using state vector propagation mode. This method becomes computationally infeasible while the simulation involves more than 20 qubits. However, TeD-Q, using tensor network contraction mode, can process the same simulation with up to 50 qubits within a comparable time.
 
\subsection{Discussion}
\label{discussion}
Generally, the state vector propagation mode of TeD-Q and PennyLane have similar performance both in CPU and GPU. This is with-in our expectation since both of them calculate the full amplitude and use backpropagation method for obtaining gradient. TeD-Q has a slight advantage ($20\sim30\%$) over Pennylane on CPU for qubits less than 11, this can be explained by the reusable mechanism that saves the time for state and gate class instantiation.

Tensor network contraction mode shows its power on large qubit systems. It can reduce the computation and memory complexity substantially. But currently its performance is constrained by inefficient use of GPU. Most pair contractions in the contraction path involve small tensors, they contribute very small portions of the overall computation and memory complexity. But the transmission of these small tensors to GPU will cost sizable overhead. This overhead problem can be improved by pre-computing contractions with small sizes on CPU and only performing the large ones on GPU\cite{huang2021efficient}, which will be included in the future release of TeD-Q.

\section{summary}
In this work, we introduced TeD-Q, an open-source software framework for quantum machine learning, variational quantum algorithm, and quantum computing simulation. With tensor contraction, simulation of quantum circuit with large number of qubits is possible. TeD-Q seamlessly integrates classical machine learning libraries with quantum simulators, allowing users to leverage the power of classical machine learning while training quantum machine learning models. TeD-Q is also empowered with the capability of training quantum machine learning models in a distributed manner. This is an important feature considering that most NISQ quantum algorithms require results from multiple instances of the same quantum circuit. In TeD-Q, the quantum circuit is seen as a Python function, which makes it easy to use. TeD-Q is device independent so that users can run the circuits on different backends, including software simulators (JAX and PyTorch) and quantum hardwares. 

To better facilitate the implementation of quantum machine learning, TeD-Q supports automatic differentiation, for instance, backpropagation, parameters shift, and finite difference methods, to obtain gradient. It also has a flexible interface that bridges quantum circuits to powerful machine learning libraries (like PyTorch). In addition, TeD-Q supports visualization functionality, where users can visualize quantum circuits and training progress in real-time.

There are still challenges in future works of TeD-Q. One crucial aspect is how to leverage distributed computation's ability to solve real quantum machine learning problems. Even though TeD-Q supports distributed computation, algorithms, and applications that make the most use of it are still lacking. Another aspect that TeD-Q can improve is to work completely compatible with different kinds of quantum hardwares on the cloud. We have demonstrated how TeD-Q can access the hardware resources on the IBM cloud in Section~\ref{section_hardware_use_case} and in later releases support for other platforms, e.g. Pennylane and Amazon, will be introduced. In principle, TeD-Q is ready to connect with any device; however, due to the lack of quantum hardware resources, this has not been tested. It would be exciting to see TeD-Q working with different hardware backends and implementing real distributed quantum algorithms, especially QML algorithms, in the future. 

\begin{acknowledgments}
We appreciate the helpful discussion with Juhua Liu.

\end{acknowledgments}

\medskip
\textbf{Author contributions.} X.-Y. W., Y.-X. D., and D.-C. T. conceived the project. Y.-C. C. and X.-Y. W. designed the architecture of the software. Y.-C. C., X.-Y. W., and C.-Y. K. did the coding and carried out the numerical simulation. X.-Y. W., Y.-X. D., Y.-C. C., C.-Y. K. and D.-C. T. analyzed the results. All authors contributed to discussions of the results and the development of the manuscript. Y.-C. C., C.-Y. K., and X.-Y. W. wrote the manuscript with input from all co-authors.  X.-Y. W., Y.-X. D., and D.-C. T. supervised the whole project.

\appendix





\nocite{*}


\bibliographystyle{quantum}
\bibliography{apssamp}

\begin{thebibliography}{10}

\bibitem{arute2019quantum}
Frank Arute, Kunal Arya, Ryan Babbush, Dave Bacon, Joseph~C Bardin, Rami
  Barends, Rupak Biswas, Sergio Boixo, Fernando~GSL Brandao, David~A Buell,
  et~al.
\newblock ``Quantum supremacy using a programmable superconducting processor''.
\newblock \href{https://dx.doi.org/10.1038/s41586-019-1666-5}{Nature {\bf 574},
  505--510}~(2019).

\bibitem{madsen2022quantum}
Lars~S Madsen, Fabian Laudenbach, Mohsen~Falamarzi Askarani, Fabien Rortais,
  Trevor Vincent, Jacob~FF Bulmer, Filippo~M Miatto, Leonhard Neuhaus, Lukas~G
  Helt, Matthew~J Collins, et~al.
\newblock ``Quantum computational advantage with a programmable photonic
  processor''.
\newblock \href{https://dx.doi.org/10.1038/s41586-022-04725-x}{Nature {\bf
  606}, 75--81}~(2022).

\bibitem{wu2021strong}
Yulin Wu, Wan-Su Bao, Sirui Cao, Fusheng Chen, Ming-Cheng Chen, Xiawei Chen,
  Tung-Hsun Chung, Hui Deng, Yajie Du, Daojin Fan, et~al.
\newblock ``Strong quantum computational advantage using a superconducting
  quantum processor''.
\newblock \href{https://dx.doi.org/10.1103/PhysRevLett.127.180501}{Physical
  review letters {\bf 127}, 180501}~(2021).

\bibitem{bergholm2018pennylane}
Ville Bergholm, Josh Izaac, Maria Schuld, Christian Gogolin, M~Sohaib Alam,
  Shahnawaz Ahmed, Juan~Miguel Arrazola, Carsten Blank, Alain Delgado, Soran
  Jahangiri, et~al.
\newblock ``Pennylane: Automatic differentiation of hybrid quantum-classical
  computations''~(2018).
\newblock  \href{http://arxiv.org/abs/2110.03402}{arXiv:2110.03402}.

\bibitem{qiskit}
A~tA~v, MD~SAJID ANIS, Abby-Mitchell, H{\'e}ctor Abraham, AduOffei, Rochisha
  Agarwal, Gabriele Agliardi, Merav Aharoni, Vishnu Ajith, Ismail~Yunus
  Akhalwaya, Gadi Aleksandrowicz, Thomas Alexander, Matthew Amy, et~al.
\newblock ``Qiskit: An open-source framework for quantum computing''.
\newblock \url{https://github.com/Qiskit/qiskit}~(2021).

\bibitem{yao}
Xiu-Zhe Luo, Jin-Guo Liu, Pan Zhang, and Lei Wang.
\newblock ``Yao. jl: Extensible, efficient framework for quantum algorithm
  design''.
\newblock \href{https://dx.doi.org/10.22331/q-2020-10-11-341}{Quantum {\bf 4},
  341}~(2020).

\bibitem{cirq}
``Cirq: A python framework for creating, editing, and invoking noisy
  intermediate scale quantum (nisq) circuits''.
\newblock \url{https://github.com/quantumlib/Cirq}~(2022).

\bibitem{biamonte2017quantum}
Jacob Biamonte, Peter Wittek, Nicola Pancotti, Patrick Rebentrost, Nathan
  Wiebe, and Seth Lloyd.
\newblock ``Quantum machine learning''.
\newblock \href{https://dx.doi.org/10.1038/nature23474}{Nature {\bf 549},
  195--202}~(2017).

\bibitem{schuld2015introduction}
Maria Schuld, Ilya Sinayskiy, and Francesco Petruccione.
\newblock ``An introduction to quantum machine learning''.
\newblock \href{https://dx.doi.org/10.1080/00107514.2014.964942}{Contemporary
  Physics {\bf 56}, 172--185}~(2015).

\bibitem{schuld2019quantum}
Maria Schuld and Nathan Killoran.
\newblock ``Quantum machine learning in feature hilbert spaces''.
\newblock \href{https://dx.doi.org/10.1103/PhysRevLett.122.040504}{Physical
  review letters {\bf 122}, 040504}~(2019).

\bibitem{huang2021power}
Hsin-Yuan Huang, Michael Broughton, Masoud Mohseni, Ryan Babbush, Sergio Boixo,
  Hartmut Neven, and Jarrod~R McClean.
\newblock ``Power of data in quantum machine learning''.
\newblock \href{https://dx.doi.org/10.1038/s41467-021-22539-9}{Nature
  communications {\bf 12}, 1--9}~(2021).

\bibitem{yuan2019theory}
Xiao Yuan, Suguru Endo, Qi~Zhao, Ying Li, and Simon~C Benjamin.
\newblock ``Theory of variational quantum simulation''.
\newblock \href{https://dx.doi.org/10.22331/q-2019-10-07-191}{Quantum {\bf 3},
  191}~(2019).

\bibitem{cerezo2021variational}
Marco Cerezo, Andrew Arrasmith, Ryan Babbush, Simon~C Benjamin, Suguru Endo,
  Keisuke Fujii, Jarrod~R McClean, Kosuke Mitarai, Xiao Yuan, Lukasz Cincio,
  et~al.
\newblock ``Variational quantum algorithms''.
\newblock \href{https://dx.doi.org/10.1038/s42254-021-00348-9}{Nature Reviews
  Physics {\bf 3}, 625--644}~(2021).

\bibitem{bittel2021training}
Lennart Bittel and Martin Kliesch.
\newblock ``Training variational quantum algorithms is np-hard''.
\newblock \href{https://dx.doi.org/10.1103/PhysRevLett.127.120502}{Physical
  Review Letters {\bf 127}, 120502}~(2021).

\bibitem{preskill2018quantum}
John Preskill.
\newblock ``Quantum computing in the nisq era and beyond''.
\newblock \href{https://dx.doi.org/10.22331/q-2018-08-06-79}{Quantum {\bf 2},
  79}~(2018).

\bibitem{bharti2022noisy}
Kishor Bharti, Alba Cervera-Lierta, Thi~Ha Kyaw, Tobias Haug, Sumner
  Alperin-Lea, Abhinav Anand, Matthias Degroote, Hermanni Heimonen, Jakob~S
  Kottmann, Tim Menke, et~al.
\newblock ``Noisy intermediate-scale quantum algorithms''.
\newblock \href{https://dx.doi.org/10.1103/RevModPhys.94.015004}{Reviews of
  Modern Physics {\bf 94}, 015004}~(2022).

\bibitem{NEURIPS2019_9015}
Adam Paszke, Sam Gross, Francisco Massa, Adam Lerer, James Bradbury, Gregory
  Chanan, Trevor Killeen, Zeming Lin, Natalia Gimelshein, Luca Antiga, Alban
  Desmaison, Andreas Kopf, Edward Yang, Zachary DeVito, Martin Raison, Alykhan
  Tejani, Sasank Chilamkurthy, Benoit Steiner, Lu~Fang, Junjie Bai, and Soumith
  Chintala.
\newblock ``Pytorch: An imperative style, high-performance deep learning
  library''.
\newblock In Advances in Neural Information Processing Systems 32.
\newblock \href{https://dx.doi.org/10.48550/arXiv.1912.01703}{Pages
  8024--8035}.
\newblock Curran Associates, Inc.~(2019).

\bibitem{aycock2003brief}
John Aycock.
\newblock ``A brief history of just-in-time''.
\newblock \href{https://dx.doi.org/10.1145/857076.857077}{ACM Computing Surveys
  (CSUR) {\bf 35}, 97--113}~(2003).

\bibitem{Paddlequantum}
``{Paddle Quantum}''.
\newblock \url{https://github.com/PaddlePaddle/Quantum}~(2020).

\bibitem{qiqcbook}
Michael~A. Nielsen and Isaac~L. Chuang.
\newblock ``Quantum computation and quantum information: 10th anniversary
  edition''.
\newblock \href{https://dx.doi.org/10.1017/CBO9780511976667}{Cambridge
  University Press}. ~(2011).

\bibitem{gray2021hyper}
Johnnie Gray and Stefanos Kourtis.
\newblock ``Hyper-optimized tensor network contraction''.
\newblock \href{https://dx.doi.org/10.22331/q-2021-03-15-410}{Quantum {\bf 5},
  410}~(2021).

\bibitem{gogate2012complete}
Vibhav Gogate and Rina Dechter.
\newblock ``A complete anytime algorithm for treewidth''~(2012).
\newblock  \href{http://arxiv.org/abs/1207.4109}{arXiv:1207.4109}.

\bibitem{daniel2018opt}
G~Daniel, Johnnie Gray, et~al.
\newblock ``Opt$\backslash$\_einsum-a python package for optimizing contraction
  order for einsum-like expressions''.
\newblock \href{https://dx.doi.org/10.21105/joss.00753}{Journal of Open Source
  Software {\bf 3}, 753}~(2018).

\bibitem{DBLP:phd/dnb/Schlag20}
Sebastian Schlag.
\newblock ``High-quality hypergraph partitioning''.
\newblock \href{https://dx.doi.org/10.1145/3529090}{PhD thesis}.
\newblock Karlsruhe Institute of Technology, Germany.
\newblock ~(2020).

\bibitem{ahss2017alenex}
Yaroslav Akhremtsev, Tobias Heuer, Peter Sanders, and Sebastian Schlag.
\newblock ``Engineering a direct \emph{k}-way hypergraph partitioning
  algorithm''.
\newblock In 19th Workshop on Algorithm Engineering and Experiments, (ALENEX
  2017).
\newblock \href{https://dx.doi.org/10.1137/1.9781611974768.3}{Pages 28--42}.
\newblock ~(2017).

\bibitem{huang2021efficient}
Cupjin Huang, Fang Zhang, Michael Newman, Xiaotong Ni, Dawei Ding, Junjie Cai,
  Xun Gao, Tenghui Wang, Feng Wu, Gengyan Zhang, et~al.
\newblock ``Efficient parallelization of tensor network contraction for
  simulating quantum computation''.
\newblock \href{https://dx.doi.org/10.1038/s43588-021-00119-7}{Nature
  Computational Science {\bf 1}, 578--587}~(2021).

\bibitem{pednault2017pareto}
Edwin Pednault, John~A Gunnels, Giacomo Nannicini, Lior Horesh, Thomas
  Magerlein, Edgar Solomonik, Erik~W Draeger, Eric~T Holland, and Robert
  Wisnieff.
\newblock ``Pareto-efficient quantum circuit simulation using tensor
  contraction deferral''~(2017).
\newblock  \href{http://arxiv.org/abs/1710.05867}{arXiv:1710.05867}.

\bibitem{paszke2017automatic}
Adam Paszke, Sam Gross, Soumith Chintala, Gregory Chanan, Edward Yang, Zachary
  DeVito, Zeming Lin, Alban Desmaison, Luca Antiga, and Adam Lerer.
\newblock ``Automatic differentiation in pytorch''.
\newblock In NIPS 2017 Workshop on Autodiff.
\newblock ~(2017).
\newblock  url:~\url{https://openreview.net/forum?id=BJJsrmfCZ}.

\bibitem{bradbury2021jax}
James Bradbury, Roy Frostig, Peter Hawkins, Matthew~James Johnson, Chris Leary,
  Dougal Maclaurin, George Necula, Adam Paszke, Jake VanderPlas, Skye
  Wanderman-Milne, et~al.
\newblock ``Jax: Autograd and xla''.
\newblock Astrophysics Source Code LibraryPages ascl--2111~(2021).
\newblock
  url:~\url{https://ui.adsabs.harvard.edu/abs/2021ascl.soft11002B/abstract}.

\bibitem{param_shift}
Maria Schuld, Ville Bergholm, Christian Gogolin, Josh Izaac, and Nathan
  Killoran.
\newblock ``Evaluating analytic gradients on quantum hardware''.
\newblock \href{https://dx.doi.org/10.1103/PhysRevA.99.032331}{Phys. Rev. A
  {\bf 99}, 032331}~(2019).

\bibitem{four_term_parameter_shift}
Gian-Luca~R Anselmetti, David Wierichs, Christian Gogolin, and Robert~M
  Parrish.
\newblock ``Local, expressive, quantum-number-preserving {VQE} ansätze for
  fermionic systems''.
\newblock \href{https://dx.doi.org/10.1088/1367-2630/ac2cb3}{New Journal of
  Physics {\bf 23}, 113010}~(2021).

\bibitem{nandkishore2015many}
Rahul Nandkishore and David~A Huse.
\newblock ``Many-body localization and thermalization in quantum statistical
  mechanics''.
\newblock
  \href{https://dx.doi.org/10.1146/annurev-conmatphys-031214-014726}{Annu. Rev.
  Condens. Matter Phys. {\bf 6}, 15--38}~(2015).

\bibitem{goodfellow2020generative}
Ian Goodfellow, Jean Pouget-Abadie, Mehdi Mirza, Bing Xu, David Warde-Farley,
  Sherjil Ozair, Aaron Courville, and Yoshua Bengio.
\newblock ``Generative adversarial networks''.
\newblock \href{https://dx.doi.org/10.48550/arXiv.1406.2661}{Communications of
  the ACM {\bf 63}, 139--144}~(2020).

\bibitem{tian2022recent}
Jinkai Tian, Xiaoyu Sun, Yuxuan Du, Shanshan Zhao, Qing Liu, Kaining Zhang, Wei
  Yi, Wanrong Huang, Chaoyue Wang, Xingyao Wu, et~al.
\newblock ``Recent advances for quantum neural networks in generative
  learning''~(2022).
\newblock  \href{http://arxiv.org/abs/2206.03066}{arXiv:2206.03066}.

\bibitem{huang2021experimental}
He-Liang Huang, Yuxuan Du, Ming Gong, Youwei Zhao, Yulin Wu, Chaoyue Wang,
  Shaowei Li, Futian Liang, Jin Lin, Yu~Xu, et~al.
\newblock ``Experimental quantum generative adversarial networks for image
  generation''.
\newblock \href{https://dx.doi.org/10.1103/PhysRevApplied.16.024051}{Physical
  Review Applied {\bf 16}, 024051}~(2021).

\bibitem{du2022quantum}
Yuxuan Du, Tao Huang, Shan You, Min-Hsiu Hsieh, and Dacheng Tao.
\newblock ``Quantum circuit architecture search for variational quantum
  algorithms''.
\newblock \href{https://dx.doi.org/10.1038/s41534-022-00570-y}{npj Quantum
  Information {\bf 8}, 1--8}~(2022).

\bibitem{linghu2022quantum}
Kehuan Linghu, Yang Qian, Ruixia Wang, Meng-Jun Hu, Zhiyuan Li, Xuegang Li,
  Huikai Xu, Jingning Zhang, Teng Ma, Peng Zhao, et~al.
\newblock ``Quantum circuit architecture search on a superconducting
  processor''~(2022).
\newblock  \href{http://arxiv.org/abs/2201.00934}{arXiv:2201.00934}.

\bibitem{mbl_theory}
Dmitry~A. Abanin, Ehud Altman, Immanuel Bloch, and Maksym Serbyn.
\newblock ``Colloquium: Many-body localization, thermalization, and
  entanglement''.
\newblock \href{https://dx.doi.org/10.1103/RevModPhys.91.021001}{Rev. Mod.
  Phys. {\bf 91}, 021001}~(2019).

\bibitem{mbl2D}
Ming Gong, He-Liang Huang, Shiyu Wang, Chu Guo, Shaowei Li, Yulin Wu, Qingling
  Zhu, Youwei Zhao, Shaojun Guo, Haoran Qian, et~al.
\newblock ``Quantum neuronal sensing of quantum many-body states on a 61-qubit
  programmable superconducting processor''~(2022).
\newblock  \href{http://arxiv.org/abs/2201.05957}{arXiv:2201.05957}.

\bibitem{huang2021}
He-Liang Huang, Yuxuan Du, Ming Gong, Youwei Zhao, Yulin Wu, Chaoyue Wang,
  Shaowei Li, Futian Liang, Jin Lin, Yu~Xu, Rui Yang, Tongliang Liu, Min-Hsiu
  Hsieh, Hui Deng, Hao Rong, Cheng-Zhi Peng, Chao-Yang Lu, Yu-Ao Chen, Dacheng
  Tao, Xiaobo Zhu, and Jian-Wei Pan.
\newblock ``Experimental quantum generative adversarial networks for image
  generation''.
\newblock \href{https://dx.doi.org/10.1103/PhysRevApplied.16.024051}{Phys. Rev.
  Applied {\bf 16}, 024051}~(2021).

\end{thebibliography}

\end{document}